\documentclass[sigconf,nonacm,screen]{acmart}

\definecolor{marroon}{HTML}{881c1c}
\definecolor{DarkBlue}{HTML}{00008B}
\definecolor{lgreen}{HTML}{e0f3db}
\definecolor{dpink}{HTML}{CD1076}
\definecolor{pink}{HTML}{FED2D2}
\definecolor{soothinggreen}{HTML}{4dac26}
\definecolor{darkred}{HTML}{8B0000}
\definecolor{dblue}{HTML}{104E8B}
\definecolor{violet}{HTML}{8A2BE2}
\definecolor{mscolor}{HTML}{01665e}
\definecolor{nmscolor}{HTML}{d8b365}
\definecolor{deepgrey}{HTML}{525252}
\definecolor{dslate}{HTML}{2F4F4F}
\definecolor{dolive}{HTML}{556B2F}
\definecolor{teal}{HTML}{388E8E}
\definecolor{gray}{HTML}{e8e8e8}

\usepackage{multirow,stfloats}
\usepackage{colortbl}
\usepackage{cleveref}
\usepackage{subcaption}
\usepackage{soul}
\soulregister\pq0
\soulregister\mq0
\soulregister\nummoderators0
\soulregister\autoref7
\soulregister\cite7
\soulregister\citet7

\newcommand{\nummoderators}{five}

\newcommand{\blockquote}[2]{%
  \begin{quote}
    \textit{``#1''} (#2)
  \end{quote}
}

\newcommand{\inlinequote}[1]{%
    \textit{``#1''}
}

\newcommand{\pq}[0]{\textsc{Positive Queue}}
\newcommand{\mq}[0]{modqueue}
\newcommand{\test}[0]{\textsc{Testbed}}

\usepackage{multirow,array,booktabs,siunitx,adjustbox}

\definecolor{chiorange}{HTML}{D55E00}

\newif{\ifshowrevisions}
  \showrevisionstrue
  \showrevisionsfalse 
\ifshowrevisions
    \newcommand{\anonymous}[1]{\textbf{\small\sffamily{\textcolor{purple}{[#1 -- Anonymous]}}}}
    \newcommand{\edit}[1]{\textcolor{blue}{#1}}
    \newcommand{\cut}[1]{\textcolor{blue}{\st{#1}}}
    \newcommand{\chiedit}[1]{\textcolor{chiorange}{#1}}
    \newcommand{\chicut}[1]{\textcolor{chiorange}{\st{#1}}\hspace{0.2em}}
    
\else
    \newcommand{\anonymous}[1]{}
    \newcommand{\edit}[1]{#1}
    \newcommand{\cut}[1]{}
    \newcommand{\chiedit}[1]{#1}
    \newcommand{\chicut}[1]{}
    
\fi

\AtBeginDocument{%
  }

\begin{document}
\sloppy

\title[Mind Your Ps and Qs: Positive Moderation Practice in the Positive Queue]{Mind Your Ps and Qs: Positive Moderation Practice in the Positive Queue}

\author{Charlotte Lambert}
\orcid{0009-0002-8487-7485}
\affiliation{%
  \institution{University of Illinois Urbana-Champaign}
  \city{Urbana}
  \state{Illinois}
  \country{USA}}
\email{cjl8@illinois.edu}

\author{Agam Goyal}
\orcid{0009-0009-5989-2887}
\affiliation{%
  \institution{University of Illinois Urbana-Champaign}
  \city{Urbana}
  \state{Illinois}
  \country{USA}}
\email{agamg2@illinois.edu}

\author{Eunice Mok}
\orcid{0009-0007-4753-3049}
\affiliation{%
  \institution{University of Illinois Urbana-Champaign}
  \city{Urbana}
  \state{Illinois}
  \country{USA}}
\email{eymok2@illinois.edu}

\author{Eshwar Chandrasekharan}
\orcid{0000-0002-7473-1418}
\affiliation{%
  \institution{University of Illinois Urbana-Champaign}
  \city{Urbana}
  \state{Illinois}
  \country{USA}}
\email{eshwar@illinois.edu}

\renewcommand{\shortauthors}{Lambert et al.}

\begin{abstract}
\edit{Online communities}\chicut{of all sizes} rely on volunteer moderators to maintain order. 
Despite their key role, moderators are given a toolbox of punishments \chicut{and asked to fend off barrages of harmful content. However, prior research shows that positive feedback may proactively encourage higher quality contributions and discourage norm violations. Moderators themselves have requested support for locating and rewarding content to encourage in their communities, yet there is a tangible lack of practical support through tools.}
\chiedit{and far less support for encouraging contributions they want to see more of.}
\chicut{We build the \textsc{Positive Queue}, a Google Chrome extension that integrates discovery, rewarding, and norm-setting features directly into Reddit's existing modqueue.}
\chiedit{We introduce the \pq{} as a positive counterpart to Reddit's modqueue: a dedicated space for moderators to discover contributions and behaviors they want to encourage and positively reinforce. With five moderators, four with 6--14 years of experience, we use the \pq{} to examine how moderators operationalize positive reinforcement.
Moderators combined predicted community reception, observed engagement, and their own judgment; used prediction--engagement mismatches to identify overlooked content; and repurposed positive features for punitive and retrospective work. These findings surface tensions around labor, attribution, and community fit. 
We contribute the \pq{} as a working system and conceptualize positive moderation as recognition infrastructure that shapes what moderators notice, whose judgment becomes visible, and how recognition reaches contributors.}
\end{abstract}

\ccsdesc[500]{Human-centered computing~Collaborative and social computing systems and tools}

\keywords{Online Communities, Community Norms, Positive Reinforcement, Recognition Infrastructure}


\maketitle

\section{Introduction}






Moderation is an ``essential, constitutional, definitional''~\cite{gillespie_custodians_2018} aspect of what platforms do. 
To platforms, careful moderation is imperative to retain users, as both too-strict and too-lenient moderation may discourage some from returning to an online space~\cite{gillespie_custodians_2018}.
To users, moderation is necessitated by the fact that ``healthy online communities have moderators who facilitate communication''~\cite{grimmelmann_virtues_2015}.

Without their valuable moderation work and efforts to establish community-wide norms, users of social media platforms face far more content posing risks to their mental health~\cite{denniss_social_2025,saha_prevalence_2019,pluta_exposure_2023}.
However, this means volunteer moderators of platforms like Reddit have aggregated hundreds of hours of work every day~\cite{li_measuring_2022}, mandating even more exposure to that same content that harms users.
On Reddit, this physically manifests in the \textit{\mq{}}, a dedicated space to perform moderation actions in which reported and removed content is funneled for manual review.
Inevitably, this moderation work on Reddit and online communities more broadly has taken a toll on moderator mental health~\cite{wohn_volunteer_2019,steiger_psychological_2021}.

Moderation is not only the practice of preventing harms, but also of facilitating cooperation and encouraging positive participation among members of a community~\cite{grimmelmann_virtues_2015}.
Preventing non-normative behavior is somewhat well-supported in that moderators across platforms have tools that discourage hate speech~\cite{chandrasekharan_you_2017} and general rule-breaking~\cite{ribeiro_automated_2022, srinivasan_content_2019} through punishments.
However, punitive actions do not teach users what they \textit{should} do, meaning moderators still need tools to encourage adoption of positive community norms and therefore accomplish the full scope of their task as moderators.




\begin{figure*}[t]
    \includegraphics[width=0.75\linewidth]
    {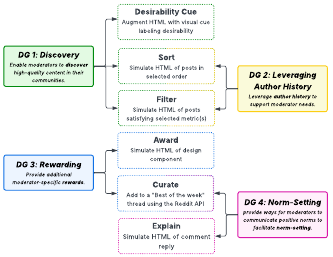}
    \caption{Schematic diagram \chicut{illustrating the four design goals of the \pq{} and the implementation of six new features added to the \mq{} via the \pq{} system to accomplish each one}\chiedit{mapping the four design goals of the \pq{} to six features added to the \mq{}}. For example, the ``Curate'' feature \chicut{enables moderators to shape user behavior by rewarding high quality content (DG~3) and setting community norms (DG~4)}\chiedit{supports rewarding high-quality content (DG~3) and communicating positive norms (DG~4)}, contrasting prior work centering curation as a means to shape feeds.}
    
    \Description{Diagram showing each design goal and arrows pointing toward the features designed to accomplish the goal. DG1 is discovery and points to the desirability cue, sorting, and filtering. DG2 is leveraging author history and points to sorting and filtering. DG3 is rewarding and points to the award and curate features. Finally, DG4 is norm-setting and points to the ``curate'' and ``explain'' features.}
    \label{fig:diagram}
\end{figure*}

\subsection{Established Need for Positive Moderation Tools}
Moderation research has begun to pivot from a sole focus on reactive, punitive moderation actions toward proactive and positive ones.
In this new space, CSCW researchers have developed tools to proactively identify and triage comments for moderator review~\cite{chandrasekharan_crossmod_2019}, encourage users to reflect before contributing to tension online~\cite{chang_thread_2022}, and curate community feeds based on predicted curator preferences~\cite{he_cura_2023}.
On the positive side, HCI theory has long touted positive feedback as a way to increase user motivation~\cite{kraut_building_2011} and enable communities to better communicate norms~\cite{kiesler_regulating_2012}\edit{, but designing tools to leverage these ideas remains underdeveloped and challenging.}

\edit{Recent research has explored} the application of \textit{positive reinforcement}---a behavioral psychology principle to encourage desired outcomes by introducing positive stimuli~\cite{ferster_schedules_1957}---to online spaces. 
Researchers established that Reddit moderators employ positive reinforcement\edit{, but doing so often requires them to} read through every post or stumble upon content in the \mq{}~\cite{lambert_positive_2024}.
These strategies for discovery are infeasible in larger communities due to the volume of content and made difficult by the \mq{}'s focus on reported content.
Furthermore, once content is discovered, moderators have to adapt existing features on the platform to provide actual rewards. 
These limitations motivated explicit requests for \edit{tools that empower the use of positive moderation strategies through explicit discovery and rewarding mechanisms.}
\chiedit{Despite limited dedicated support for these positive moderation practices,}
recent research showed that positive feedback, even given out anonymously through upvotes, encourages recipients to contribute higher-quality content and violate community norms less frequently~\cite{lambert_does_2025}\edit{, potentially reducing moderator workload in the long run}. 
This supports positive reinforcement as a mechanism for change capable of encouraging the positive community norms that punitive moderation actions cannot.

\subsection{Introducing the \pq{}}




\edit{We contribute the \pq{}, a \chicut{novel }\textit{positive moderation} system implemented as a Google Chrome browser extension.}
\chiedit{We use the \pq{} to study how experienced moderators operationalize positive reinforcement when explicit positive moderation affordances are available.}
\chicut{The \pq{} allows moderators to engage with positive reinforcement strategies not currently supported explicitly through Reddit's native moderation tools.} \chiedit{The \pq{} serves as a positive counterpart to the \mq{}: whereas the \mq{} surfaces content that may require intervention, the \pq{} provides a dedicated place to discover and reinforce content and behaviors moderators want to encourage.}
The \pq{} relies on the fact that moderators already use positive reinforcement without dedicated ways to do so from their platforms and have requested new design interventions to support them~\cite{lambert_positive_2024}.
Furthermore, the use of positive feedback has been shown to causally improve user level outcomes (e.g., norm-adherence and content quality)~\cite{lambert_does_2025}.
\chicut{Therefore the \pq{} has the potential to satisfy moderator needs and support adoption of an effective method of moderation leveraging positive feedback.}\chiedit{Together, this evidence motivates explicit support for positive reinforcement in moderator workflows.}

The system is built as a Google Chrome browser extension that adds to Reddit's \mq{}~\cite{bajpai_think_2025}, to minimize disruptions to existing workflows for both positive (e.g., flairs, highlights) and punitive (e.g., removals, bans) moderation strategies, while introducing new features to enable the use of positive reinforcement.
\chiedit{Existing mechanisms such as Flair and Highlights can reward or showcase content once a moderator encounters it, but they do not provide a dedicated workflow for finding contributions to encourage. The \pq{} connects that discovery problem to positive action in one moderation space.}
\chiedit{Within that space, the design choice doing the work is not the presence of reward mechanisms, several of which already exist on Reddit, but how contributions are surfaced for positive reinforcement: predicting how a community may receive a contribution rather than ranking only what it has already rewarded.}

While prior work often focuses on detecting, triaging, or visualizing content and
conversations that require moderator attention~\cite{liu_needling_2025,choi_convex_2023,bagga_are_2021,zhang_conversations_2018},
the \pq{} both facilitates detection and equips moderators with positive interventions
in the form of rewards and explanations.
The design consists of organizational features to aid discovery of high quality content (e.g., sorting, filtering), AI-based visual cues to draw moderator attention to content predicted to \chicut{be desirable}\chiedit{receive strong community reception}, new mechanisms for rewarding high-quality content, and a \chicut{novel }\chiedit{dedicated }way to communicate positive community norms via explanations. 
\autoref{fig:diagram} maps these design goals to the system features they motivate and \autoref{fig:overall} shows the full system.

\chiedit{This study does not evaluate whether repeated use of the \pq{} changes community-level norms; prior causal work motivates positive reinforcement as an intervention. Instead, we examine how experienced moderators operationalize positive reinforcement when explicit support for discovery and action is available.}
\chiedit{Accordingly, our evaluation asks how moderators use and reason about this positive moderation space.
We ask two research questions:}

\chiedit{\textbf{RQ1.} How do experienced moderators combine system-provided signals, realized community engagement, and their own judgment when deciding which contributions to encourage?}

\chiedit{\textbf{RQ2.} What tensions, appropriations, and broader moderation practices become visible when positive reinforcement affordances are integrated into existing moderation workflows?}

\subsection{Summary of Contributions}
\edit{This work makes three contributions to the underdeveloped space of positive moderation tools.
\chicut{First, we contribute the \pq{}, a Google Chrome extension that integrates discovery, rewarding, and norm-setting features---ones previously explored individually but not together---directly into Reddit's existing moderation space, the \mq{}. This addresses a crucial gap in positive moderation tools explicitly identified in a recent study.}
\chiedit{First, we provide an empirical account of how experienced moderators operationalize positive reinforcement when explicit positive moderation affordances are available. With five experienced moderators---four with 6--14 years of experience---we show how they combine predicted community reception, observed engagement, and their own judgment; identify overlooked content through mismatches among these signals; and appropriate positive features for punitive and retrospective work.}
\chicut{Second, we use the \pq{} to surface insights about how moderators interact with, reason about, and repurpose positive moderation features through a user study with \nummoderators{} experienced moderators. We find that moderators adapted the \pq{} for several unexpected workflows including finding undervalued content, connecting to the members of the community, and even performing punitive moderation tasks.}
\chiedit{Second, we synthesize these findings into a view of positive moderation as \textbf{recognition infrastructure}: systems that shape what moderators notice, whose judgment becomes visible, and how recognition reaches contributors. We derive design considerations around community reception, attribution and channel, labor, configurability, and the porous boundary between positive and punitive moderation.}
\chicut{Third, we synthesize these patterns into design considerations for future positive moderation tools across platforms---such as viewing attribution and channel of positive feedback as independent design axes, supporting interactions between features, embracing community variety through configurability, and emphasizing retrospective and auditing mechanisms to elicit a community strategist mindset.}
\chiedit{Third, we contribute the \pq{}, a working system that instantiates a dedicated positive moderation space within Reddit's existing workflow by connecting discovery with rewarding and norm-setting actions.}}

\section{Background}

This section summarizes the body of relevant work in social computing and behavioral psychology. Additional background literature is described in \autoref{sec:design-goals} to motivate the design goals underpinning the \pq{}.

\subsection{The Reddit \textit{Modqueue}}

The Reddit platform provides its volunteer moderators with a dedicated moderation interface: the \textit{\mq{}}. 
The \mq{} contains curated feeds of unmoderated, reported, flagged, and removed content from the communities a user moderates. 
On these feeds, moderators can apply \textit{actions} to posts and comments, primarily ``Approve'' and ``Remove''. 
The interface also allows actions such as marking spam, locking threads to prevent additional engagement, adding flairs to posts, and adding posts to the community highlights.
The \mq{} contains reordering \textit{mechanisms}, such as sorting and filtering, aimed at surfacing content needing moderator attention.

\textit{The \pq{} augments the current \mq{}, an interface for moderators to apply punitive actions, to define a \chicut{novel}\chiedit{dedicated} space for moderators to apply rewarding actions while maintaining support for all existing features.}

\subsection{Proactive vs. Reactive Moderation}

Reactive  approaches have long dominated the conversation around content moderation. 
Researchers have measured the effectiveness of punitive interventions~\cite{chandrasekharan_quarantined_2022,chandrasekharan_you_2017,ribeiro_automated_2022,srinivasan_content_2019, jhaver_does_2019}, proactively detected content to moderate~\cite{bagga_are_2021,davidson_automated_2017,waseem_hateful_2016,zhang_conversations_2018,mariconti__2019}, and developed new punitive strategies for moderation~\cite{jhaver_designing_2022,mahar_squadbox_2018,chang_thread_2022,im_synthesized_2020}.
One thing common amongst nearly all punitive moderation interventions is the inability to teach users how they \textit{should} behave in a community.
Removal explanations---a Reddit feature letting moderators reply to removed Reddit posts explaining the rules violated---are one example of a reactive moderation approach with some capacity to educate users about norms~\cite{jhaver_does_2019}.
While these explanations may help the recipient avoid repeating the violation, they are opaque to other community members who may not have seen the removed content prior to its removal. As a result, these explanations do not decrease onlookers' rate of removal in the community~\cite{jhaver_bystanders_2024}. 
Furthermore, removal explanations only occur when something has gone wrong and Reddit does not have support for explanations when something goes right.

As an alternative, research increasingly advocates for utilizing proactive strategies in moderation which  can enable better application of punitive measures, by
detecting negative conversational or community-level outcomes~\cite{zhang_conversations_2018,habib_act_2019} and preemptively intervening to keep conversations on track~\cite{schluger_proactive_2022}. 
On the other hand, proactive moderation can also involve the identification of conversations with prosocial outcomes~\cite{lambert_conversational_2022}, the use of
non-punitive strategies like moderator- and community-given rewards~\cite{lambert_positive_2024, lambert_does_2025} to teach users how they should behave and prevent future norm violations, or encouraging positive participation through example~\cite{seering_shaping_2017}.

\textit{The \pq{} supplements existing reactive moderation features on the \mq{} with \chicut{novel}\chiedit{dedicated} ways to help moderators identify places to apply proactive moderation through positive actions.}

\subsection{Behavioral Psychology Principles to Leverage in Proactive Moderation Techniques}

Principles from behavioral psychology can guide researchers of online governance toward proactive moderation strategies.
One example is \textit{positive reinforcement}, the practice of introducing a desirable stimulus in response to desired behavior to increase the likelihood of that behavior continuing in the future~\cite{ferster_schedules_1957}.
Positive reinforcement has been validated as an effective strategy for encouraging behaviors in varying offline contexts~\cite{armstrong_gamification_2018,perryer_enhancing_2016,boyd_direct_1981,steinberg_impact_1992,weinberg_effect_1990,wiese_sport_1991}
and researchers have proposed applying the principle to encourage desired behaviors online~\cite{lambert_positive_2024}.
\citet{kraut_building_2011} highlight positive feedback as capable of encouraging users to participate.
\citet{kiesler_regulating_2012} note the importance of highlighting norm-adhering behavior to
influence bystanders to repeat those behaviors (i.e., \textit{vicarious reinforcement}).
Quantitatively, \citet{wang_highlighting_2021} show the power of a New York Times badge in encouraging higher quality and more frequent contributions. 
Similarly, \citet{lambert_does_2025} demonstrate how receiving forms of positive feedback on Reddit causes improvements in content quality, increases in posting frequency, and reduced rates of norm violations.
Finally, researchers have revealed the power of gamified interface features to encourage user behaviors, such as Stack Overflow badges' ability to encourage participation~\cite{anderson_steering_2013}.

\textit{This work builds upon principles in behavioral psychology and leverages the established causal power of positive feedback on Reddit to develop 
\edit{a moderator-facing system that integrates positive reinforcement into the modqueue interface where Reddit moderators already work}.}

\subsection{Existing Space of Moderation Tools}
Moderation research has often explored system design as a way to make practical interventions and elicit insights into how moderators perform their tasks. This includes tools designed to 
help YouTube moderators create word filters for automated moderation~\cite{jhaver_designing_2022}, 
provide feedback to users in situations that do not warrant removals or bans~\cite{seering_chillbot_2024}, 
facilitate apologies after conflict~\cite{doan_design_2025}, 
and locate toxic discussions easily through visualization~\cite{choi_convex_2023, liu_needling_2025}.

Given the importance of removing harmful content, many of these  moderation tools  focus on applying punitive measures, though some HCI system-builders have developed tools empowering the use of proactive approaches.
These include tools for moderation teams, such as 
ModSandbox~\cite{song_modsandbox_2023}, which helps moderators visualize the outcomes of various automated rules and proactively improve their implementation.
There are also user-focused tools in this space.
\citet{chang_thread_2022} built a tool to encourage users to self-moderate as they draft comments by reporting live feedback when their drafted comment would escalate tension in a conversation.
Similarly, \citet{horta_ribeiro_post_2025} introduced a post guidance feature to actively guide users toward community guidelines as they draft posts, shown effective at increasing norm-abiding posts and lessening moderator workload.
\citet{mahar_squadbox_2018} develop Squadbox, a collaborative tool that distributes the role of moderating emails to a user's friends.

\edit{However, coupled with platforms' focus on punitive actions, designing for positive moderation strategies like positive reinforcement is difficult to operationalize, leaving the need partially unfulfilled. 
Some challenges include the wide variance between norms and values across communities~\cite{fiesler_reddit_2018, lambert_positive_2024, kriplean_articulations_2008, chandrasekharan_internets_2018, goyal_uncovering_2024, horne_identifying_2017}, the abundance of ways moderators currently moderate~\cite{bajpai_think_2025, seering_who_2023, cullen_practicing_2022, dosono_moderation_2019,kiene_technological_2019, wohn_volunteer_2019, seering_moderator_2019}, and user-level differences dictating different socialization and feedback needs~\cite{lambert_does_2025,choi_socialization_2010,morgan_evaluating_2018, halfaker_snuggle_2014}.}

Despite \edit{these challenges}, there is certainly interest in using design to support positive reinforcement in moderation~\cite{lambert_positive_2024}. 
Despite developing a visualization tool to help highlight areas of toxicity, \citet{liu_needling_2025} reveal that moderators are interested in tools that incorporate new techniques for positive reinforcement.
\edit{Some existing tools partially support positive moderation strategies, such as CommentIQ, which aids news outlet moderators in proactively finding high quality comments~\cite{park_supporting_2016}.
Similarly, Cura~\cite{he_cura_2023} leverages a model to detect content likely to align with  designated curator preferences to shape a community's front-facing feed.
However, these tools do not fully address Reddit moderators' need for a centralized moderation tool to not only find, but also reward high-quality content in their communities~\cite{lambert_positive_2024}.}

\textit{The \pq{} contributes to this \edit{underdeveloped} space of \edit{positive moderation} tools and fills the gap identified in prior work by  giving moderators agency to discover content they find desirable and to  intervene to shape user behaviors through positive reinforcement}.

\subsection{Moderator Perspectives on Automated Tools}

Given the rich space of automated moderation tools, it is important to consider moderator perspectives when developing new ones. 
Research in this space has surfaced valuable insights, namely that moderators value agency and the ability to personalize their automated tools.
\citet{atreja_appealmod_2024} emphasize that despite challenges reconciling automation with agency, moderators prioritized retaining power over final moderation decisions when using tools. 
There is also evidence that users value agency in the context of personalized moderation, such as hiding ableist hate speech~\cite{heung_ignorance_2025}.

In addition to the emphasis on agency, other work reveals that some moderators do not wish to impose their own views on what deserves to be rewarded and instead want the community's own preferences to be highlighted~\cite{lambert_positive_2024}.
Given this prior work, we ensure that the \pq{} leaves it to moderators to make final decisions about what should be rewarded, if anything. In addition, we incorporate AI-driven metrics alongside author statistics and other native Reddit metrics that moderators can use as they wish.

\chicut{By using community signals, such as upvotes, as primary metrics to discover high-quality
content, the \pq{} enables moderators to further highlight contributions their communities value.}

\textit{The \pq{} prioritizes empowering moderators to make well-informed decisions based on factors they value while ensuring agency over all final decisions.}

\section{Design Goals for Enabling Positive Moderation Practice}
\label{sec:design-goals}

This section presents the four design goals used to instantiate a positive-moderation space in the \pq{} (visualized in \autoref{fig:diagram}). Grounded in prior work on moderation tools and positive reinforcement, these goals define the affordances made available to moderators. The subsequent study examines the practices that become visible as moderators use, combine, and appropriate these affordances.
\chicut{Each objective was synthesized from a review of literature related to moderation tools and perspectives on positive reinforcement. Relevant background literature is presented as motivation for each design objective.}

\subsection{Discovery of High Quality Content}

To effectively enable positive reinforcement strategies in moderation, moderators must be able to find content they want to encourage.
Prior work shows that norm-violations often go unmoderated~\cite{park_measuring_2022,chandrasekharan_crossmod_2019}, suggesting that high-quality content may similarly go un-rewarded.
Existing methods of locating content to encourage include
reading through all the content posted in their communities, browsing posts normally, or sifting through the \mq{}~\cite{lambert_positive_2024}.
However, Reddit moderators often spend more than 5 hours per week moderating, and many spend more than 20~\cite{li_measuring_2022,bajpai_think_2025}.
Furthermore, the threaded structure of discussions in the \mq{} present challenges in parsing manually~\cite{liu_needling_2025}.
\chicut{Thus, current manual methods may not be suited for many communities---especially large ones---and leave ample room for high-quality contributions to go unnoticed by moderation teams. Moderators can use sorting and filtering to reduce the amount of content to search through.}\chiedit{These constraints leave room for contributions moderators may want to encourage to go unnoticed.} However, \citet{bajpai_think_2025} found that the sorting and filtering mechanisms on the \mq{} have minimal options and significant functionality issues, therefore not meeting moderator needs. 
Beyond improvements to sorting and filtering options, moderators have requested explicit support for discovering high-quality content specifically~\cite{lambert_positive_2024}.

\begin{quote}
   \textit{\textbf{Design Goal 1.} The \pq{} should enable moderators to \textbf{discover} high-quality content in their communities.} 
\end{quote}

\subsection{Leveraging Author History}
Among requests for better mechanisms for organizing and refining the \mq{}, moderators have articulated the need for more integration of author history in their moderation tools. 
Some moderators utilize author history in their existing workflows~\cite{bajpai_think_2025}, often by leaving the \mq{} interface to retrieve relevant information.
Furthermore, evaluations of third-party tools indicate that moderators across platforms need author history to make informed moderation decisions~\cite{liu_needling_2025,choi_convex_2023}.
Thus, the minimal integration of author history in the \mq{} limits their ability to identify content aligning with their values. 
\chicut{Other tools that have included user history in their design demonstrate how author history can empower moderators to make decisions appropriate for each user.}

Along with broader interest in author history to inform moderation decisions, some explicitly highlight support for newcomers as a core value in their moderation~\cite{lambert_positive_2024}.
\chicut{Social media research similarly highlights the importance of supporting newcomers, a group that requires special attention to retain and socialize effectively. Quantitative research has examined what environments encourage newcomers to participate, how they adapt to community norms, how long they remain active, and how automated moderation treats new contributors. This line of research elucidates the particular vulnerability of newcomers and motivates work on how communities support them. Researchers have also shown that feedback can encourage newcomers to share, newcomer badges can discourage unwelcoming reactions, and positive feedback has particularly strong impacts on newcomers' future behavior. Because this group is especially susceptible to the effects of moderation, communities would benefit from tools that consider author age.}
\chiedit{Newcomers warrant particular attention because early experiences shape whether they participate, adapt to community norms, and remain active~\cite{kraut_challenges_2011,seering_proximate_2020,rajadesingan_quick_2020,danescu-niculescu-mizil_no_2013}. Feedback and newcomer-specific design can help, and positive feedback has especially strong effects on newcomers' future behavior~\cite{burke_feed_2009,santos_can_2020,lambert_does_2025}.}

\edit{Thus, author history serves positive moderation in two ways: it helps moderators support newcomers, whose early contributions benefit most from encouragement, and identify established contributors worth recognizing or recruiting to their mod teams.}

\begin{quote}
\textit{\textbf{Design Goal 2.} The \pq{} should \textbf{leverage an author's history} to better support moderator needs, newcomer integration, and the safety of the community.} 
\end{quote}

\subsection{Rewarding of High-Quality Content}
Beyond discovering high-quality content, we aim to provide interventions that moderators can use to shape behavior. These constitute the actual rewarding mechanisms driving positive reinforcement.
In their survey, \citet{lambert_positive_2024} revealed that Reddit lacks moderator-specific ways to reward content and moderators use workarounds to adapt existing mechanisms (e.g., community sidebar, comments) to encourage desired behaviors. 
Even though surveyed moderators had access to features like flairs and stickying, 
these features typically have other primary purposes (e.g., organization) and moderators expressed their limited ability to serve as rewards.
Thus, many explicitly requested features with more targeted support for positive reinforcement. 
Researchers have shown that providing additional reward mechanisms may have real-world impact~\cite{cunha_warm_2017}, encourage higher-quality content~\cite{lambert_does_2025}, discourage norm violations~\cite{lambert_does_2025}, and maintain user motivation~\cite{kraut_building_2011}.
Furthermore, feedback from authority figures in a community has been shown to impact engagement and other desirable outcomes~\cite{seering_shaping_2017, choi_creator_2024}.
Despite their potential to positively influence communities as the de facto authority figures, Reddit moderators are discouraged from rewarding high-quality contributions both because the platform has no explicit ways to do so and because there is no precedent in their moderation team~\cite{lambert_positive_2024}.
By augmenting the \mq{} with rewarding mechanisms, platforms can implicitly communicate that providing positive feedback is a valid form of moderation. 

\begin{quote}
\textit{\textbf{Design Goal 3.} The \pq{} should support existing ways moderators \textbf{reward} high-quality content and provide additional mechanisms explicitly intended for positive moderator feedback.} 
\end{quote}

\subsection{Norm-Setting Positive Norms}
In alignment with the goals of positive reinforcement, users should be aware of what behavior was rewarded to understand how to continue contributing successfully to a community.
This connects to norm-setting, the idea of instilling norms in a community, defined as one of four main techniques of moderation and deemed ``moderation's biggest challenge and most important mission''~\cite{grimmelmann_virtues_2015}.
In practice, \citet{grimmelmann_virtues_2015} specifies that moderators can clearly define a community's norms by articulating them, for example through explanations of moderation decisions.
\citet{kiesler_regulating_2012} also claim that displaying formal feedback to community members can help others learn and comply with norms.
Prior work shows removal explanations on Reddit have the power to reduce the odds that a user's future posts will be removed~\cite{jhaver_does_2019} 
and post guidance may increase norm-abiding posts and improve moderator workload~\cite{horta_ribeiro_post_2025}.
Beyond proactive and reactive explanations of negative moderation decisions, \citet{jurgens_just_2019} assert that communities can adopt a ``capabilities approach'' which involves communicating norms in terms of what users \textit{can} do.
Although there is still research to be done to understand whether positive explanations can quantifiably encourage norm-setting, we argue it is important that moderators have the ability to provide them.

\begin{quote}
   \textit{\textbf{Design Goal 4.} The \pq{} should allow moderators to communicate the attributes of a contribution they want to encourage to facilitate \textbf{norm-setting} in their communities.} 
\end{quote}

\section{Developing the \pq{} as a Browser Extension}

\begin{figure*}[t]
    \centering
        \includegraphics[width=\linewidth]{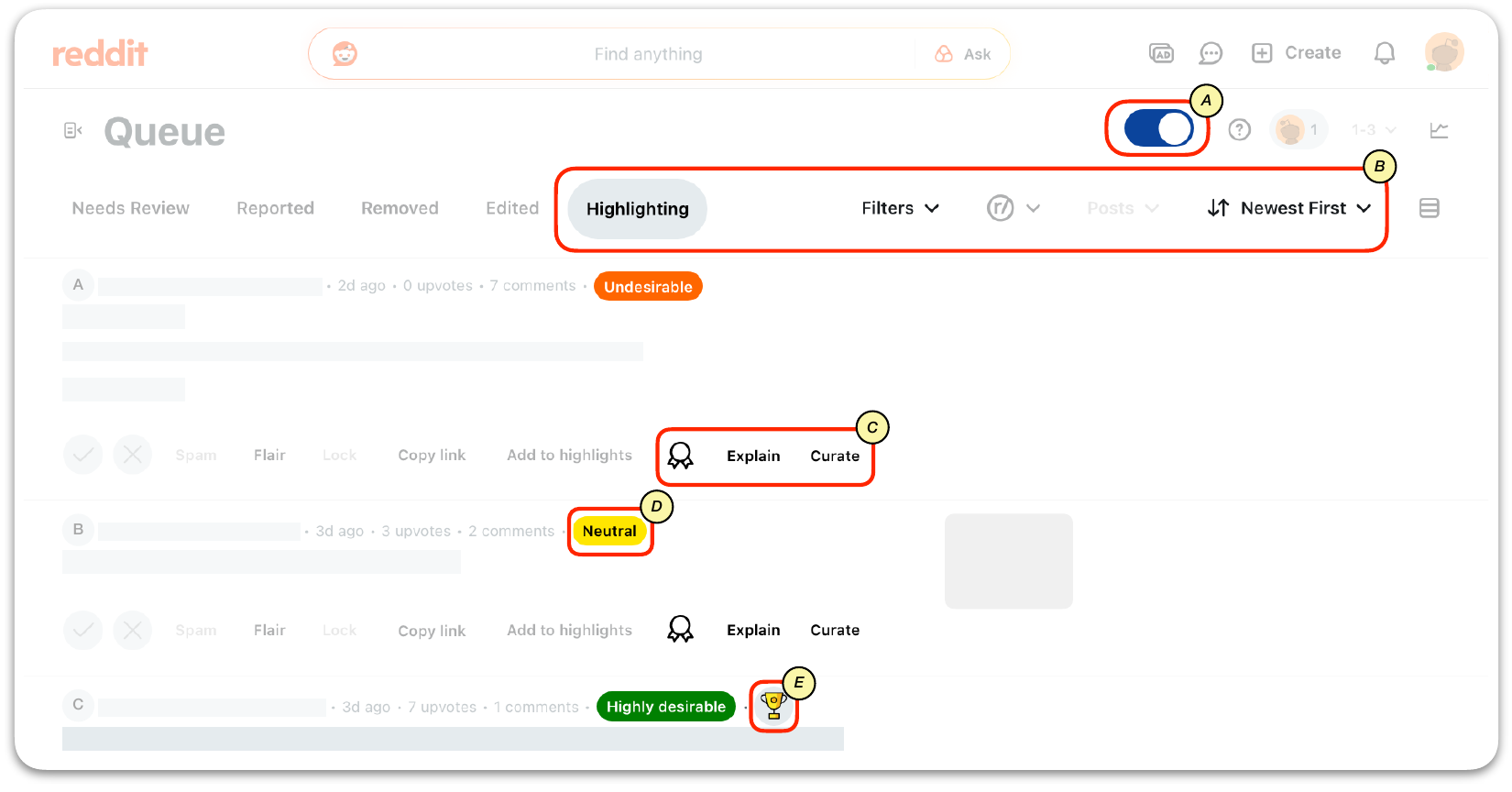}
    \caption{Overall design of the \pq{} including newly implemented features that overlay the existing \mq{} enabled by a switch (A). Once enabled, the browser extension augments the \textit{filter bar} (B) by renaming the ``Unmoderated'' feed to ``Highlighting'', inserting a new filter menu, and adding new sorting options. The system supplements each post's \textit{action row} (C) with new actions (award, ``Explain,'' and ``Curate''), and each post is enriched with a visual cue (D) denoting the post's \chicut{expected desirability}\chiedit{predicted community reception, labeled as desirability in the evaluated interface}. Awarded posts are also adorned with a trophy icon (E). All features can be removed by turning the switch (A) off.}
    \Description{Screenshot of the Positive Queue turned on overlaying the modqueue containing three visible, but anonymized posts. There are red boxes highlighting each region of the system and labeled (A) through (E) as described in the caption.}
    \label{fig:overall}
\end{figure*}

\begin{figure*}[t]
    \centering
        \includegraphics[width=0.85\linewidth]{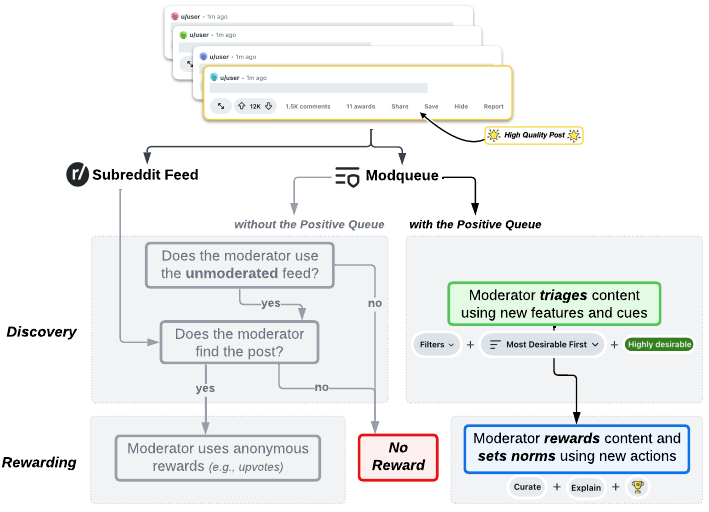}
    \caption{Two example workflows for discovering and rewarding high-quality content on Reddit. \chicut{The left of the diagram demonstrates how a moderator must find and reward content without the \pq{}, namely by searching through the entire community feed or all content in the \mq{}. Without the \pq{}, there are many opportunities for content to not be found and additional rewarding challenges. On the right, the workflow to discover and reward content is bolstered by the presence of the \pq{}.}\chiedit{The left shows discovery and rewarding using existing Reddit interfaces; the right shows the additional discovery and rewarding actions available through the \pq{}.}}
    \Description{Flowchart demonstrating workflows with and without the positive queue. The diagram starts with a selection of posts, one of which is labeled "high quality". Then the diagram follows workflows, either ending in a rewards through anonymous means, rewards through new actions from the \pq{}, or no reward at all.}
    \label{fig:workflow}
\end{figure*}

\edit{The \pq{} builds on the stated design goals (\autoref{sec:design-goals}) 
as a browser extension that augments Reddit's existing modqueue interface.}
\chicut{This section describes the design and implementation of the \textit{\pq{}} as a dedicated space for moderators to find high-quality content and administer rewarding actions.}
\chiedit{This section describes the system's design and implementation.}
\chiedit{\autoref{fig:workflow} contrasts discovery-and-rewarding workflows using existing Reddit interfaces with the additional actions available through the \pq{}.}
\edit{\chicut{It also enables us to elicit moderator thoughts on how best to incorporate design for positive moderation into existing moderation spaces.}}
\chicut{As visualized in \autoref{fig:overall}, the features of the \pq{} are added on top of the existing \mq{} interface. Each feature of the \pq{} is mapped to the design goal(s) it aims to accomplish in \autoref{fig:diagram}. \autoref{fig:workflow} shows existing moderator workflows to discover and reward high quality content compared against the workflow afforded by the \pq{}.}

\subsection{Tool Format}

The \pq{} is developed as a Google Chrome extension to integrate new features into the existing Reddit moderator interface (the \textit{\mq{}}) while minimizing disruptions to existing workflows. 
HCI researchers have previously employed browser extensions on social media to deescalate tense discussions on Reddit~\cite{chang_thread_2022}, inform Twitter users about other users' history of toxicity and misinformation~\cite{im_synthesized_2020}, gain insight into user perspectives on algorithmic timelines~\cite{wang_lower_2024}, and keep users aware of the accuracy of information online~\cite{jahanbakhsh_browser_2024}.
The tool specifically applies to the \mq{}'s ``Unmoderated'' queue, a collection of posts that have not been removed or reported, since these are the most likely posts to be reward-worthy.
As seen in \autoref{fig:overall}, the users of the \pq{} can still take any existing moderation actions, including ``Flair'' and ``Add to highlights,'' two actions moderators previously reported using for positive reinforcement~\cite{lambert_positive_2024}.

\subsection{UI Features for Discovery}
We implement three new features to facilitate the discovery of high-quality content (Design Goal 1) by helping moderators surface content deserving of rewards: a desirability visual cue, enhanced filtering, and enhanced sorting. 

\begin{figure}[t]
    \centering
        \centering
        \includegraphics[width=.7\linewidth]{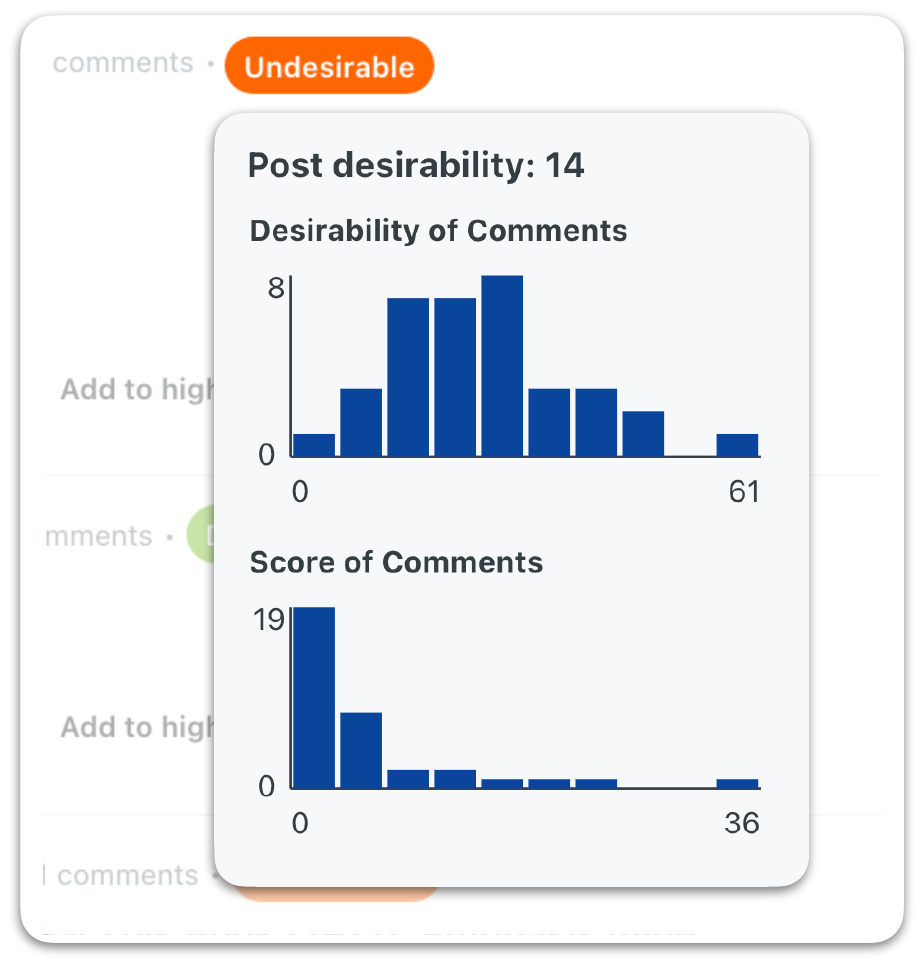}
    \Description{Panel showing that the current post is labeled ``Undesirable'' because its desirability is 14. There is a histogram of desirability in the comment section showing a near normal distribution with comments in the range 0 to 61 for desirability. The histogram of comment scores shows most comments with low score (i.e., 0-2) and a few outlier comments reaching scores up to 36.}
    \caption{This figure shows the panel that appears when a moderator hovers over the desirability cue on a post. The hover panel reports the post's \chicut{desirability score}\chiedit{predicted community reception using the desirability label shown to participants} and presents visualizations of the desirability scores in the comment section.}
    \label{fig:desirability_hover}
\end{figure}
\subsubsection{\chicut{Desirability Visual Cue}\chiedit{Predicted Community Reception}}
The first \chicut{novel }feature of the \pq{} is a visual cue denoting \edit{a contribution's \chicut{\textit{desirability}}\chiedit{predicted \textit{community reception}}.}
\chiedit{The evaluated interface labeled this cue \textit{desirability} (\autoref{fig:overall}, \autoref{fig:desirability_hover}); throughout the paper, we use \textit{predicted community reception} for the underlying construct: the likelihood that a contribution will receive a high score in its subreddit.}
\edit{We ground this notion in prior work showing that community \textit{scores} on Reddit (an aggregation of a contribution's upvotes and downvotes) can serve as a proxy for what kind of content a community values and finds desirable~\cite{goyal_uncovering_2024,goyal_language_2025}, and that moderators
actively seek to encourage desirable contributions in their communities via upvotes~\cite{lambert_positive_2024}\chicut{, and that receiving positive feedback via upvotes leads to higher-quality contributions}.
\chicut{These prior studies tie score to the behavior moderators want to reinforce rather than to popularity.}
\chiedit{Recent work likewise uses community-specific reception to characterize community values and approval patterns~\cite{goyal_language_2025,goyal_vastu_2026}. We use causal evidence about the effects of receiving upvotes~\cite{lambert_does_2025} only to motivate positive reinforcement as an intervention.}
\chiedit{Because a moderator's upvote is aggregated with other community votes, whole-community reception is the observable signal available for the upvoting behavior moderators report using to reinforce content they value~\cite{lambert_positive_2024}. Our study examines when this broader signal aligns with moderator judgment.}
\chiedit{Furthermore, Reddit itself uses score to assign user reputation (i.e., karma). Score also drives platform-wide surfacing of popular content through the r/popular feed~\cite{chan_understanding_2024}.}
Accordingly, we operationalize \chicut{desirability}\chiedit{this cue} as a contribution's predicted \textit{community reception}---its likelihood of receiving a high score in its subreddit---and surface it as an AI-driven signal in the \mq{}.
Similar to our usage of karma and score, we use \chicut{this notion of desirability}\chiedit{predicted community reception} as a feature and examine how moderators use it for positive moderation.}
\chiedit{Crucially, the cue predicts rather than observes reception. Potentially popular Reddit content can go overlooked when first submitted~\cite{gilbert_widespread_2013}; predicted reception can therefore surface contributions before engagement is fully realized and, through prediction--engagement mismatches, identify ``under-valued gems'' that may benefit from moderator attention (\autoref{sec:hidden-gems}).}
\edit{Because definitions of what different communities value or find desirable vary~\cite{weld_what_2022,weld_making_2024,park-etal-2024-valuescope,kumar-etal-2025-compo}, we adopt a community-centric approach to modeling in which we train separate models for each community using their own score signal, rather than imposing a single cross-community notion of desirability.} 
\chiedit{This choice is supported by evidence that community-specific models outperform global ones on this task~\cite{goyal_vastu_2026}.}
We opted for a trained model grounded in actual community feedback (i.e., Reddit scores) over alternatives such as randomized or heuristic-based scores, which could undermine the ecological validity of the study environment and confuse participating moderators encountering scores misaligned with their experience governing their communities. 
\edit{This is one principled operationalization of the desirability cue;} the \pq{}'s architecture is intentionally model-agnostic, so more expressive approaches---including classifiers trained on moderator actions, and community-specific language models~\cite{goyal_vastu_2026}---can be substituted in as communities identify what best captures their notion of desirability, without changing the rest of the \pq{}. \edit{Our findings (\autoref{sec:discovery-feedback}) point to concrete ways future versions could enrich this signal.}

Following the modeling methodology of \citet{goyal_language_2025}, we trained subreddit-specific models for each community recruited  (see \autoref{sec:recruitment}) using data from Pushshift archives~\cite{baumgartner_pushshift_2020} aligned with the date range used by \citet{lambert_does_2025}. We compute four sets of features \edit{grounded in these attributes}:

\begin{enumerate}
    \item[(i)] \textit{Linguistic Inquiry and Word Count 2015 (LIWC2015)~\cite{pennebaker2001linguistic}:} Following prior psychology and HCI research~\cite{fast_empath_2016,lester_content_2010,bahgat_liwc_2022,zhou_synthetic_2023}, we use the LIWC dictionary to calculate the proportion of words aligning with psycholinguistic categories such as affect, positive/negative emotion, cognitive processes, and function words. 
    \item[(ii)] \textit{Surface-level attributes:} We assess sentiment using the VADER compound score~\cite{hutto2014vader}, which ranges from $-1$ (strongly negative) to $1$ (strongly positive). We also measure readability of the post or comment with Flesch’s formula~\cite{kincaid_derivation_1975}, capture interrogative style through the proportion of sentences ending in a question mark, and evaluate content politeness using ConvoKit~\cite{chang2020convokit}.
    \item[(iii)] \textit{Toxicity:} We use an open-source toxicity model, Detoxify~\cite{hanu_detoxify_2020}, to score post and comment toxicity ranging from $0$ (non-toxic) to $1$ (extremely toxic).
    \item[(iv)] \textit{Sentence Embeddings:} We also use the SentenceBERT~\cite{reimers_sentence-bert_2019} model \texttt{all-MiniLM-L6-v2} to compute a 384-dimensional embedding of each post and comment to complete the set of features. 
\end{enumerate}

We use Reddit \textit{score}---an aggregation of the number of upvotes and downvotes---\edit{to denote community reception},
an approach used in prior work~\chiedit{\cite{goyal_uncovering_2024,goyal_language_2025,goyal_vastu_2026}}.
We bin the score of every post or comment into quartiles, then
assign the binary label \textit{``desirable''} to the top-most quartile and \textit{``undesirable''} to the bottom two quartiles. As in prior work~\cite{goyal_uncovering_2024,lambert_does_2025}, we discard the third quartile to maintain a buffer between the two categories. 

We then use subreddit-specific XGBoost models~\cite{chen2016xgboost} with \texttt{max\_depth=6} separately for posts and comments, using a $80{:}20$ train--test split, to model the likelihood of a particular post or comment \chicut{being ``desirable''}\chiedit{receiving strong community reception} using the set of computed features and embeddings. \chiedit{XGBoost is a standard strong baseline for tabular prediction, where tree ensembles remain competitive with or better than deep models~\cite{shwartz2022tabular,grinsztajn2022tree}.} 
\chiedit{We selected XGBoost over $\ell_2$-regularized logistic regression trained on the same features, which performed worse on held-out data.}
\chicut{As a comparison, we also trained $\ell_2$-regularized logistic regression models on the same features and found that XGBoost consistently achieved higher accuracy and AUC on held-out data, so we used XGBoost as part of the \chicut{desirability model}\chiedit{community-reception model}.}

\begin{table}[t]
\small
\sffamily
\centering
\caption{Subreddit-specific test accuracy and AUC for XGBoost classifiers that predict whether a post or comment would be in the top quartile for Reddit score. We report results for each model trained for subreddits we used during our user study and mean performance across models.} Models are trained separately for posts and comments with an $80{:}20$ train--test split. 

\Description{Table of accuracy scores for each subreddit that ended up in our set of interviewed moderators. We report the accuracy and AUC scores for posts and comments individually for each subreddit.}
\label{tab:subreddit-accuracy-auc}
\begin{adjustbox}{max width=0.75\textwidth}
\resizebox{\columnwidth}{!}{
\begin{tabular}{crccrcc}
\toprule
\multirow{2}{*}{\textbf{Subreddit}} & 
\multicolumn{3}{c}{\textbf{Posts}} & \multicolumn{3}{c}{\textbf{Comments}} \\
\cmidrule(lr){2-4}\cmidrule(lr){5-7}
 & \textbf{Dataset size}
 & \textbf{Accuracy} & \textbf{AUC} & \textbf{Dataset size}
 & \textbf{Accuracy} & \textbf{AUC} \\
\midrule
$\mathcal{S}_1$ & 6,467 &  76.5\% & 0.727 & 18,422 & 72.5\% & 0.608   \\ 
$\mathcal{S}_2$ & 8,791 &  77.8\% & 0.749  & 10,152& 73.4\% & 0.616  \\ 
$\mathcal{S}_3$ & 11,024 &  78.3\% & 0.780 & 79,812 & 74.9\% & 0.584  \\ 
\cmidrule(lr){3-4} \cmidrule(lr){6-7}
Mean &  & 77.5\% & 0.752 & &73.6\% & 0.603 \\
\bottomrule
\end{tabular}}
\end{adjustbox}
\end{table}

\autoref{tab:subreddit-accuracy-auc} reports subreddit-specific performance metrics and means across models.
\chicut{The model performance is reasonable for this setting as predicting whether a contribution will fall in the top quartile of score based on content alone is noisy and an inherently difficult task.} 
\chicut{The goal of the model is to provide a reasonable signal of a contribution's expected reception from the community---to help moderators understand and give feedback on the cue--rather than to maximize predictive accuracy.}\chiedit{The reception model is a noisy decision signal rather than ground truth for contribution quality. Comment-level performance is modest in particular, which limits how confidently the cue should be interpreted in isolation. Our empirical claim instead concerns how moderators interpret and combine such a signal with observed engagement and their own judgment.}

Given the model's estimated \chicut{desirability scores}\chiedit{community-reception scores}, we add a visual cue to each post and comment colored and labeled with one of five \chiedit{interface }desirability categories based on percentiles: 
``Highly desirable'' ($> 80^\text{th}$ percentile), ``Desirable'' (61$^\text{st}$--80$^\text{th}$ percentile), ``Neutral'' (41$^\text{st}$--60$^\text{th}$ percentile), ``Undesirable'' (21$^\text{st}$--40$^\text{th}$ percentile), and ``Highly undesirable'' ($\leq 20^\text{th}$ percentile). 
\chiedit{These labels are, in retrospect, a limitation of the evaluated design: they render a descriptive prediction about reception in normative language, and a moderator reading ``Highly undesirable'' may take it as a judgment about quality or appropriateness that the model does not support. A reception-framed vocabulary, such as ``likely to be well received,'' would better match the construct, and we recommend it for future implementations.}
\autoref{fig:overall} shows how the queue of posts is augmented with desirability cues on each post.

Hovering over the desirability cue reveals the contribution's desirability score. For posts, moderators can also see visualizations of the comment section activity (see \autoref{fig:desirability_hover}) including histograms of the desirability scores and Reddit scores. The visualizations are inspired by other moderation tools~\cite{choi_convex_2023,liu_needling_2025} \chicut{and may help moderators quickly identify whether a comment section has comments worth rewarding}\chiedit{and summarize comment-section reception and score distributions at a glance}.

\subsubsection{Enhanced Filtering Options to Re-Organize the Queue}

Currently, Reddit allows moderators to filter the \mq{} based on subreddit (for moderators of multiple communities), content type (i.e., text post, comment, etc.), and other relevant attributes (e.g., marked as spam, has reports, has a flair). However, these native filtering options are not visible on the feed by default.
To aid the discovery of high-quality content (i.e., Design Goal 1),
the \pq{}'s filtering menu is always visible and is positioned near Reddit's other organization menus (i.e., subreddit selection, sorting, etc.) for ease of use by experienced moderators.

\begin{figure}[t]
    \centering
    \begin{subfigure}[b]{0.3\linewidth}
        \centering
        \includegraphics[width=\linewidth]{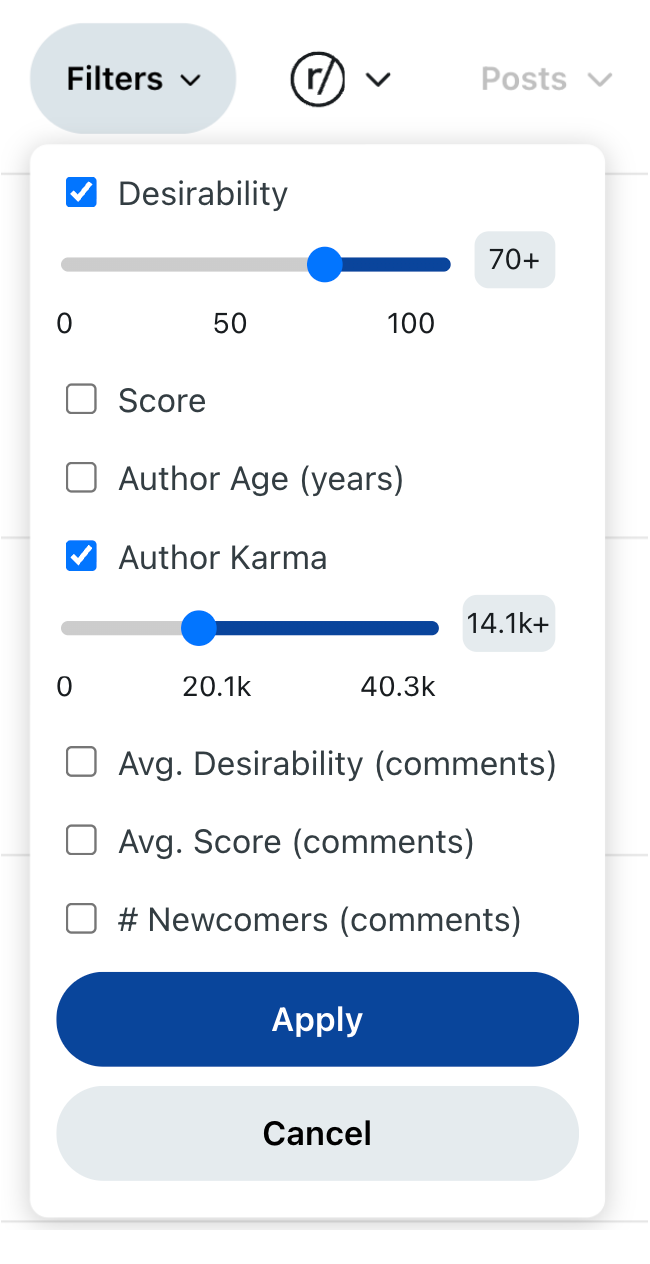}
        \caption{Filter menu} 
        \Description{Dropdown menu showing filtering options. Each option has a checkbox and when checked, a slider appears beneath the metric which a user can adjust. The figure shows two filters selected: desirability which is set to 70 and author karma which is set to 17.2k.}
        \label{fig:filter-menu}
    \end{subfigure}
    \hspace{1cm}
    \begin{subfigure}[b]{0.3\linewidth}
        \centering
        \includegraphics[width=\linewidth]{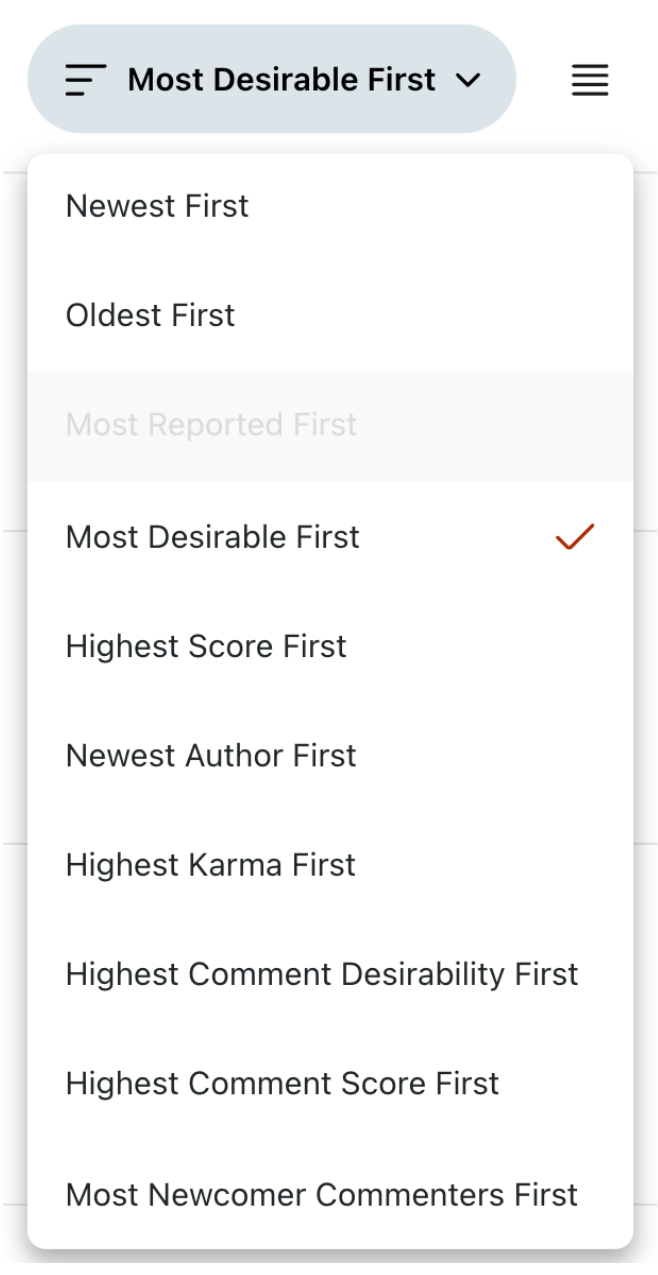}
        \caption{Sorting menu} 
        \Description{Dropdown menu showing sorting options. These options are: Newest First, Oldest First, Most Reported First, Most Desirable First, Highest Score First, Newest Author First, Highest Karma First, Highest Comment Desirability First, Highest Comment Score First, and Most Newcomer Commenters First. The Most Reported First option is grayed out because it is disabled for the purpose of the study. There is a checkmark icon on the Most Desirable First option to indicate that the queue is currently being sorted based on most desirable first.}
        \label{fig:sorting-menu}
    \end{subfigure}
    \caption{(a) Filtering and (b) sorting  menus on the \mq{}. The filter menu is \chicut{completely novel menu}\chiedit{a new menu} allowing moderators to select filters to narrow down what appears in their feed. The sorting menu existed in Reddit's existing \mq{} with only the first three sorting options. The remaining options were added by the \pq{}.}
    \label{fig:filter_sorting}
\end{figure}

More specifically, we introduce seven filtering options (shown in \autoref{fig:filter-menu}) that reduce the visible posts in the queue to the potentially high-quality ones.
The options are either at the post (desirability and score), author (author age and karma, i.e., a Reddit-specific reputation score), or comment section level (average desirability, average score, and number of newcomers). 
Each filtering attribute is one moderators already use or requested to help them identify desirable content~\cite{lambert_positive_2024}. 
These options allow filtering based on community reception, aspects of the actual author (shown to impact the effect size of the positive feedback~\cite{lambert_does_2025}), and comment-level features to enable locating desirable comments. 
The author-specific metrics target Design Goal 2 empowering moderators to leverage author-level information (i.e., age and karma) to discover high-quality content.

\chicut{Moderators can use the checkboxes to apply as many filtering metrics as they would like, allowing for granular filtering. Each selected metric has an associated slider moderators can use to select the minimum value of the metric that they want to see in the queue. That threshold appears in text to the right of the slider. The desirability metric only allows values between 0 and 100 (to mimic the model output). The remaining metrics all have a minimum value of 0 and a maximum determined dynamically based on the top 20th percentile of that metric in the given subreddit. Aside from author age, the scales are restricted to have whole number steps to be intuitive to users. The posts that satisfy all the selected filter conditions are made visible and the rest are hidden from view.}\chiedit{Moderators can combine filters and set thresholds to narrow the queue.}

\subsubsection{Enhanced Sorting Options to Re-Organize the Queue}
\chicut{oAlongside new filtering options, we augment the existing sorting menu with the same seven metrics introduced for filtering. \autoref{fig:sorting-menu} shows the augmented sorting menu where the first three metrics are Reddit-provided. The ``Most Reported First'' option is disabled for the study because the study environment has no reports. When a new option is selected, all the posts satisfying any selected filters are re-ordered accordingly. By adding options to the existing sorting menu, there are no disruptions to moderators' current understanding and process of how to sort the queue. Similar to the added filtering options, the enhanced sorting addresses both Design Goals 1 and 2 by leveraging author features, among others, to aid moderators in discovering desirable content.}\chiedit{We add the same seven metrics to Reddit's sorting menu (\autoref{fig:sorting-menu}).}

\subsection{UI Features for Rewarding}
Beyond discovery of content, we address Design Goal 3 with a new moderator-specific reward mechanism (``Curate'') and  improve support for the existing actions they take (e.g., upvote, award, flair, highlight).

\subsubsection{Curate}
\begin{figure}[t]
    \centering
    \begin{subfigure}[b]{0.55\linewidth}
        \centering
        \includegraphics[width=\linewidth]{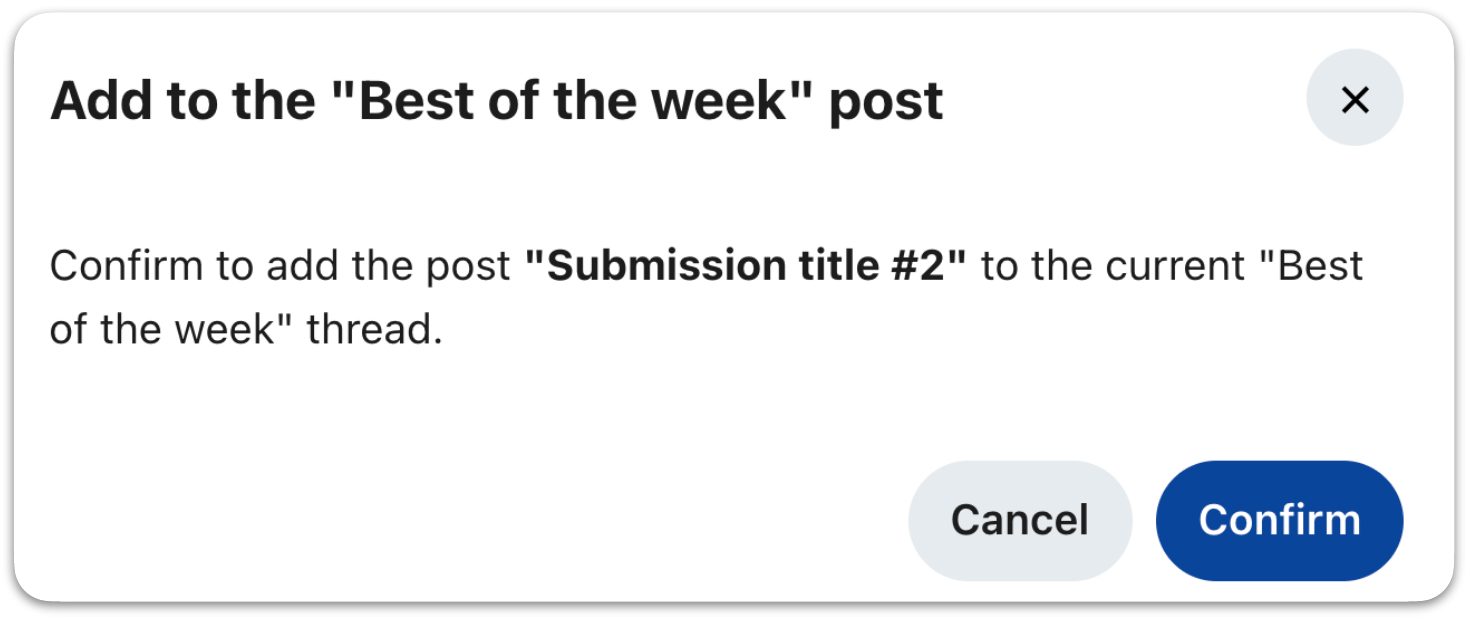}
        \caption{Confirmation pop-up prompting moderators to add a post to the ``Best of the week'' post.} 
        \Description{Screenshot of a pop-up asking for confirmation to add a sample post entitled ``Submission title #2'' to the current thread.}
        \label{fig:curate_confirm}
    \end{subfigure}
    \begin{subfigure}[b]{0.9\linewidth}
        \centering
        \includegraphics[width=\linewidth]{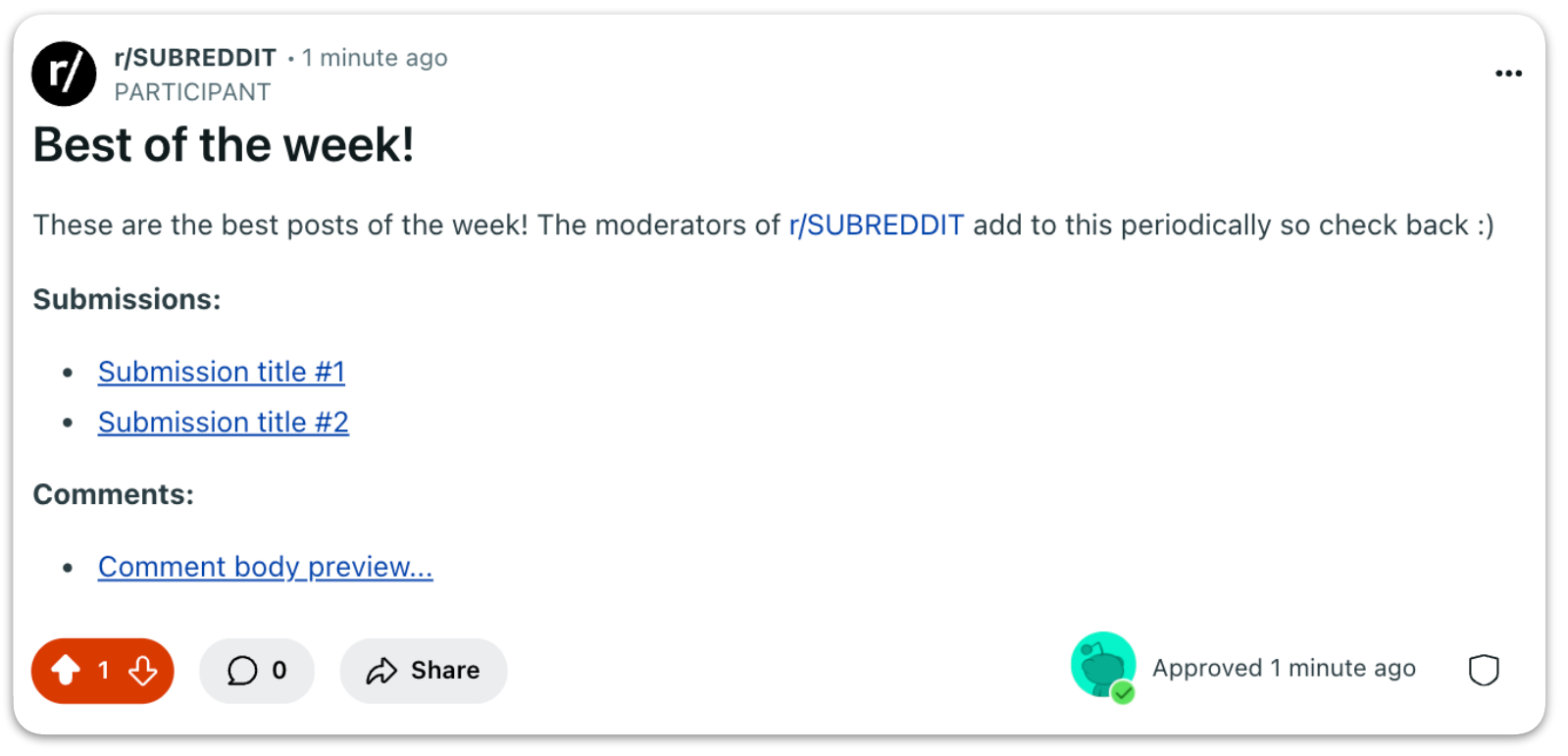}
        \caption{Reddit post created after the curate feature is used in (a) to add a new post titled ``Submission title \#2'' to the ``Best of the week'' post. A link to the post is added in the ``Submissions'' section of the thread. The post is automatically approved by the moderator's account so it does not appear in the \mq{}.} 
        \Description{Screenshot of a Reddit post entitled ``Best of the week.'' The post lists two submission titles hyperlinked and one comment body also hyperlinked. The screenshot shows that the post was approved by a moderator.}
        \label{fig:curate_post}
    \end{subfigure}
    \caption{Interface elements associated with the ``Curate'' feature. In both (a) and (b), template values are used for anonymity.}
    \label{fig:curate}
\end{figure}

The \chicut{novel}\chiedit{main new} rewarding feature in the \pq{} is the ``Curate'' button which enables moderators to quickly compile a list of high-quality posts or comments.
\chicut{r/AskHistorians, a popular Q\&A community on Reddit, currently creates weekly digests to highlight interesting, overlooked, and high-quality posts and comments. Similarly, r/help, the Reddit tech support community, posts a weekly recap highlighting helpful users and sometimes top posts.}\chiedit{Some communities already compile recurring digests of high-quality or overlooked contributions.}
Some moderators also compile links to high quality posts and comments in the sidebar of a community~\cite{lambert_positive_2024}.
\chicut{Even though moderators already compile these collections of positive examples, it is not explicitly supported by Reddit's moderation interface and thus requires significant manual effort to maintain. In 2024, Reddit introduced the ``Add to Highlights'' feature which allows moderators to pin six posts to the top of a community's feed.}
\chicut{This feature}\chiedit{Reddit's ``Add to Highlights''} addresses previous moderator complaints about only being able to pin two posts~\cite{lambert_positive_2024}, but still provides no way to highlight comments or create multiple, permanent curated lists.
\chicut{The ``Curate'' feature is intended to facilitate existing curation practices by streamlining the manual process. The feature also allows moderators to highlight both posts \textit{and} comments and structures the curation on a weekly basis, similar to existing community practices.}\chiedit{Curate collects both posts and comments into a weekly thread.}
\chiedit{\autoref{fig:curate} shows the confirmation step and resulting curated thread.}
Curation systems do exist in prior work~\cite{he_cura_2023}, but primarily focus on shaping feeds \edit{and do not necessarily empower curators to highlight content that has explicitly \textit{not} gained engagement (similar to r/AskHistorians' weekly thread highlighting overlooked posts)}.

\chicut{To contrast, the ``Curate'' feature shapes user behavior by using highlighting as a reward (Design Goal 3) and educating the community on their norms through positive example (Design Goal 4).}\chiedit{The ``Curate'' feature uses highlighting as a reward (Design Goal 3) while making moderator-selected examples visible to the community (Design Goal 4).}

\chicut{For convenience, the ``Curate'' button is located along the \textit{action row} at the bottom of each post and comment (see \autoref{fig:overall}). When the ``Curate'' button is clicked, moderators are prompted to confirm their decision (see \autoref{fig:curate_confirm}) before adding the post's title or comment's body to the corresponding section in the ``Best of the week'' thread (see \autoref{fig:curate_post}).}

\subsubsection{Support for existing rewards}

\begin{figure}[t]
    \centering
    \begin{subfigure}[b]{0.75\linewidth}
        \centering
        \includegraphics[width=\linewidth]{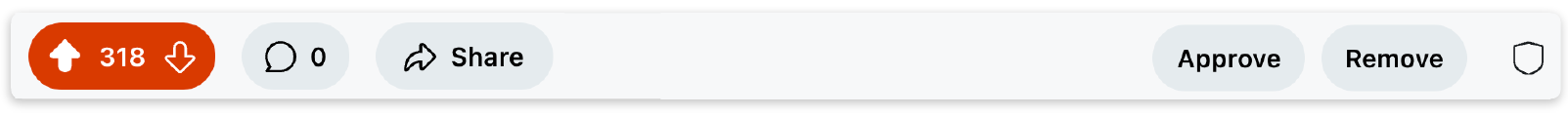}
        \caption{Built in Reddit actions on a post.} 
        \Description{Card view actions available on a Reddit post before the positive queue is enabled.}
        \label{fig:old_actions}
    \end{subfigure}
    \begin{subfigure}[b]{0.75\linewidth}
        \centering
        \includegraphics[width=\linewidth]{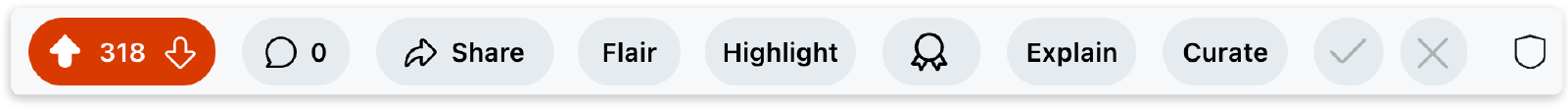}
        \caption{Actions on a post in Reddit's card view after enabling the \pq{}. The ``Flair'' and ``Highlight'' buttons are Reddit's own actions \chicut{made more easily accessible}\chiedit{surfaced in the action bar}, while the award button (represented by the medal icon), the ``Explain'' button, and the ``Curate'' button are new additions from the browser extension.} 
        \Description{Card view actions available on a Reddit post after the positive queue is enabled.}
        \label{fig:new_actions}
    \end{subfigure}
    \caption{The action bar on each post before and after enabling the \pq{} plugin.}
    \label{fig:action_bar}
\end{figure}

The \pq{} also maintains support for features moderators already use to positively reinforce content~\cite{lambert_positive_2024}. 
\chicut{In addition, Reddit's \mq{} has two different views, compact view and card view, which allow the same moderation actions, though card view obscures some common forms of positive feedback behind a menu. \autoref{fig:old_actions} shows the actions available for each post in card view along with the shield icon indicating a moderator menu containing the option to add a flair or to add the post to the community's ``highlights''---a carousel of up to six moderator-selected posts pinned to top of a subreddit. Moderators use both features as rewards, but their location in a menu introduces friction that may discourage others widespread usage. To alleviate this friction, the \pq{} adds both features directly to the action bar, as shown in \autoref{fig:new_actions}. These buttons connect to Reddit's existing workflow for handling these tasks, so the behavior is unchanged.}\chiedit{The \pq{} also moves Reddit's existing Flair and Highlight actions into the action bar (\autoref{fig:new_actions}) without changing their behavior.}

Moderators also use awards to encourage desirable behavior~\cite{lambert_positive_2024}, a feature only available to sufficiently large subreddits. 
Because the user study takes place within a small, private subreddit---so as not to manipulate real communities during the user study---we did not have access to Reddit's awards and therefore simulated them (see \autoref{sec:testbed}).
One notable difference is that the simulated awards are intentionally unlimited and free, while Reddit's are paid. 
This addresses requests for unlimited awards by moderators surveyed by \citet{lambert_positive_2024}.
The simulated awards appear next to the \chicut{novel }desirability cue, as shown in box (E) of \autoref{fig:overall}.

\subsection{UI Features for Norm-Setting}

For Design Goal 4, we implement an ``Explain'' button that gives moderators a way to communicate positive norms. Although rewards on their own can signal what a community values or considers desirable, explanations highlight which attributes are being rewarded---making the norm explicit instead of leaving it to be inferred.
\subsubsection{Explain}
\begin{figure}[t]
    \centering
    \begin{subfigure}[c]{0.5\linewidth}
        \centering
        \includegraphics[width=\linewidth]{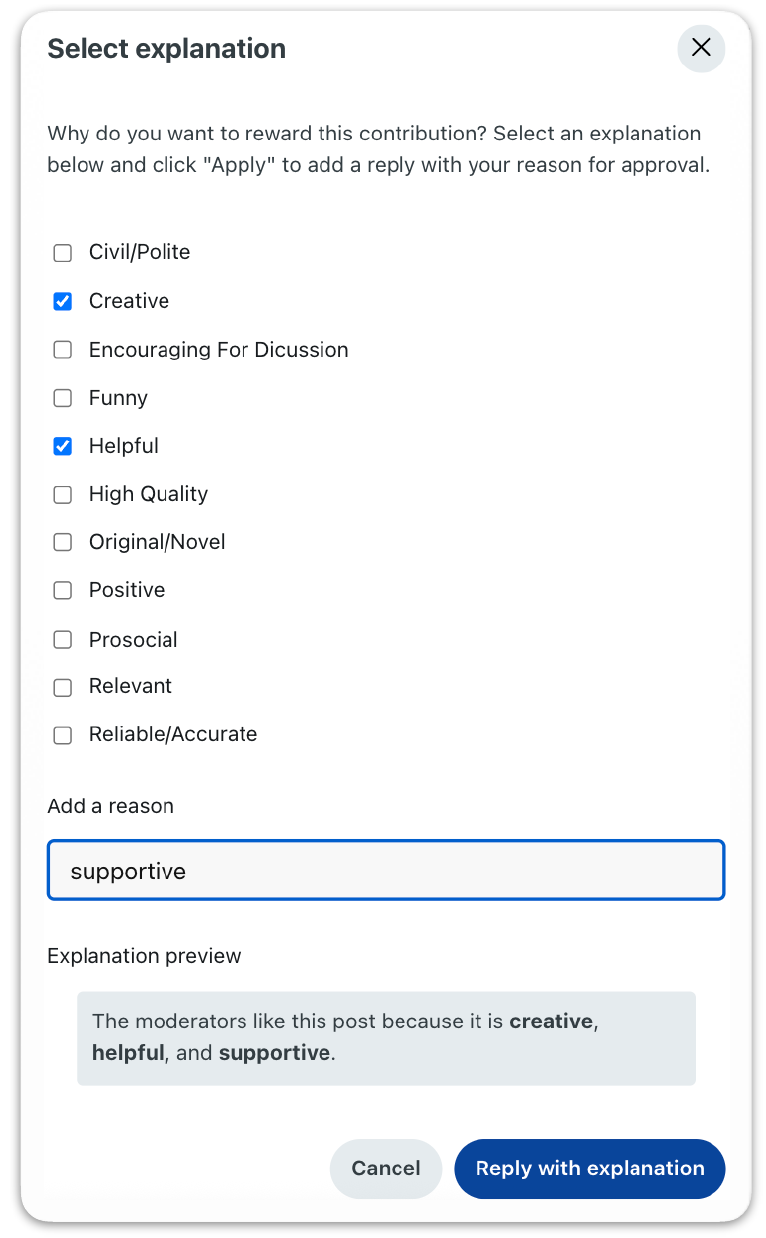}
        \caption{Explanation selection panel.} 
    \Description{The explanation window prompting users to select reasons why a contribution is high quality. There are eleven options, two of which are selected (``Creative'' and ``Helpful''). There is a custom reason box in which ``supportive'' has been typed. The pop-up has an explanation generated from these three reasons: ``The moderators like this post because it is creative, helpful, and supportive.''}
        \label{fig:explain-popup}
    \end{subfigure}
    \hspace{1cm}
    \begin{subfigure}[c]{0.8\linewidth}
        \centering
        \includegraphics[width=\linewidth]{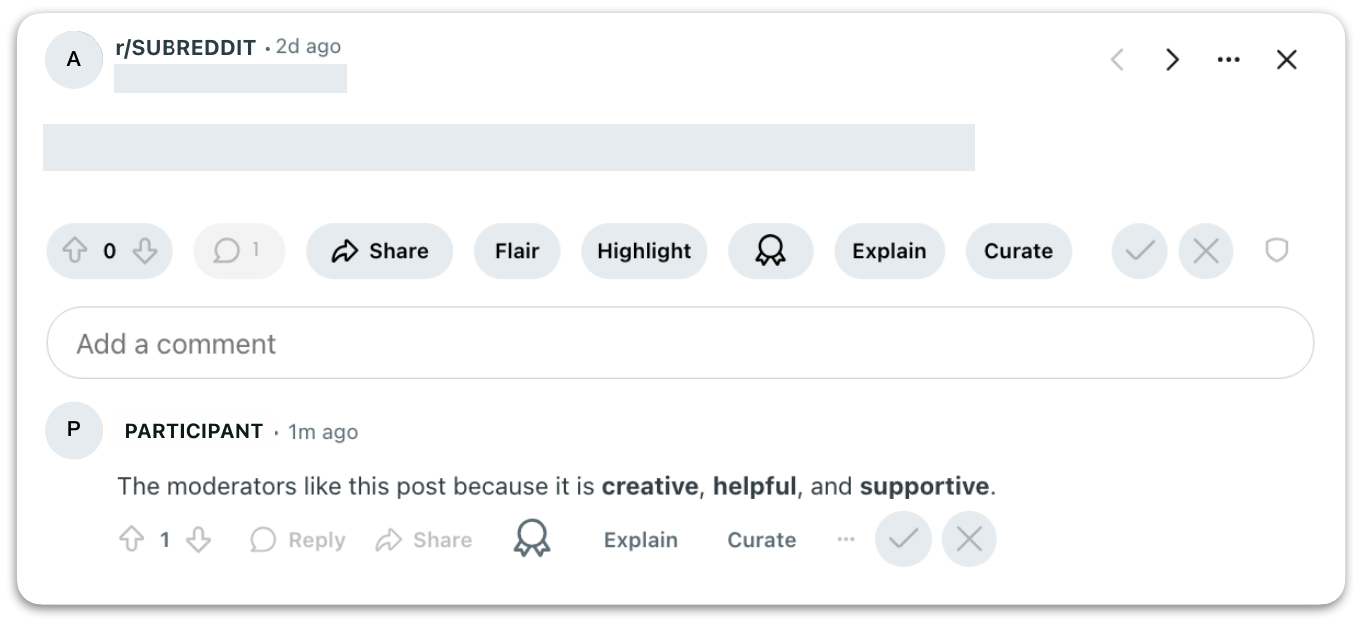}
        \caption{Explanation reply following selection from (a).}
        \Description{Reply by an anonymized "PARTICIPANT" to an anonymized post on Reddit. The reply states "the moderators like this post because it is creative, helpful, and supportive."}
        \label{fig:explain-comment}
    \end{subfigure}
    \caption{Pop-up allowing moderators to select why something might be desirable given default options (drawn from the taxonomy of desirability from \citet{lambert_positive_2024} (a). Following selection, the explanation is added as a reply to the post or comment (b).\vspace{-8pt}}
    \label{fig:explain}
\end{figure}

\chicut{Each post and comment is augmented with a new ``Explain'' button as shown in \autoref{fig:overall} and \autoref{fig:action_bar}. Similar to the ``Curate'' button, the ``Explain'' button is added to the action row to match Reddit's interface style. A button click triggers the pop-up window shown in \autoref{fig:explain-popup} which includes a prompt to select reasons why the corresponding contribution is high-quality.}\chiedit{The ``Explain'' action lets moderators select why a contribution is high-quality and preview a reply (\autoref{fig:explain-popup}).}
The pop-up provides eleven pre-populated reasons moderators have previously reported finding content desirable~\cite{lambert_positive_2024} along with an input box for a custom reason to account for variance in community values~\cite{weld_making_2024, goyal_uncovering_2024}.
\chicut{Custom reasons will persist if an ``Explain'' button is clicked again, allowing moderators to adapt the feature for their preferences and communities. The preview at the bottom of the pop-up shows a live explanation based on selected and typed reasons. Clicking ``Reply with explanation'' triggers JavaScript connecting to the Reddit API to automatically reply to the selected post or comment with the previewed explanation, as seen in \autoref{fig:explain-comment}.}

This feature is a structured way moderators can explicitly communicate their reasons for approving of a contribution to help with norm-setting~\cite{grimmelmann_virtues_2015, kiesler_regulating_2012}. 
\chicut{Because the explanation appears as a public reply, the explanation communicates positive norms to the recipient along with any bystanders, maximizing the potential for norm-acquisition through  targeted positive and vicarious reinforcement.}\chiedit{Because the explanation appears as a public reply, it makes moderator reasoning visible to the recipient and bystanders, providing a mechanism for communicating positive norms.}
It also simplifies responding to a desirable contribution, which 50\% of moderators surveyed by \citet{lambert_positive_2024} \chiedit{reported doing.}

\subsection{Piloting the System with HCI Researchers}

\chiedit{We piloted the study with four internally recruited HCI researchers; their feedback led us to replace a numeric red-to-green reception score with the five categorical labels evaluated in the study, as well as to add a Curate confirmation step and preview Explain replies before posting.}
\chicut{We conducted two mock study sessions with four internally-recruited participants to collect feedback on the \pq{}, catch lingering bugs, and ensure that the plan for the user study was reasonable.}
\chicut{After each session, we incorporated participant suggestions to improve the tool's design and functionality.}
\edit{\chicut{Previously, the desirability cues were labeled with a numeric score on a red-to-green gradient; pilot participants suggested higher-level categories to make them easier to parse.}}
\chicut{Participants also suggested the confirmation pop-up as seen in \autoref{fig:curate_confirm} to the ``Curate'' functionality. Finally, we added the explanation preview to the ``Explain'' pop-up \autoref{fig:explain-popup} to help moderators know ahead of time what actual text would be sent as a reply.}

\subsection{Artifact Availability}
\chiedit{Upon publication, we will release the source code for the \pq{} system, including the browser extension and the code used to train and evaluate the community-reception models, to support replication and future development of positive moderation tools.}

\section{\edit{User Study}}
\edit{
\chicut{We conduct a user study with active Reddit moderators that serves two complementary purposes. First, we evaluate how well the implementation of the \pq{} accomplishes the stated design goals and gather moderator feedback to improve the system's design. Second, we use the \pq{} to surface emergent patterns: what new workflows the system enables, which features moderators rely on, and how they combine affordances in unanticipated ways.}
\chiedit{We conduct a user study with experienced Reddit moderators to examine how they operationalize positive reinforcement while using the \pq{}, while also gathering feature-level feedback. We distinguish task-scaffolded behavior from unanticipated appropriations and participant-proposed uses.}
}
This section details participant recruitment, the study environment, 
and finally the user study.

\subsection{Recruiting Moderators}
\label{sec:recruitment}

Recruiting participants for the user study was done in \edit{four} waves.
We aimed to be cognizant of the increasing demand on moderators to participate in research such as ours. 
\chiedit{Direct modmail outreach is established in Reddit moderation research, including systems studies~\cite{song_modsandbox_2023,koshy_venire_2025,liu_needling_2025} and recent work on the modqueue~\cite{bajpai_think_2025}. Following this practice, }we reached out to moderators manually---through Reddit's modmail system---and gradually over time.
Additionally, we targeted communities we expected would be interested in adopting a tool for positive moderation strategies. This included communities already engaging in such strategies despite limited resources, those that often engage with HCI researchers, and those who previously indicated interest~\cite{liu_needling_2025,lambert_positive_2024}.

Each community received a recruitment message detailing the purpose of the study, the eligibility criteria (i.e., 18 years or older, fluent in English, located in the United States, and have access to a computer with an internet connection and Google Chrome), and other relevant information, along with a link to a short eligibility screening survey.
The screening survey verifies that participants meet the inclusion criteria and obtains the necessary contact information to schedule an interview. 

\chiedit{Across four waves, we contacted moderators of 10, 4, 14, and 21 candidate subreddits, respectively. The first and third waves yielded three and two participants, the second and fourth yielded none, although one moderator in the fourth initially expressed interest but did not schedule an interview.}

Ultimately, recruitment reached \edit{49} moderation teams consisting of \edit{more than 2,000} unique moderators and yielding \nummoderators{} participants, a response rate of about 0.25\%.
\autoref{tab:participants} reports information about each of the subreddits moderated by the participants---including the topic and number of subscribers at the time of this writing---and each participant's moderation experience. 

The sample of participants primarily includes highly experienced moderators of 10 distinct, varying-sized communities ranging from less than one thousand subscribers to more than 4 million and spanning 9 unique topics.
\chicut{We deemed this sample sufficient to perform an exploratory evaluation of the \pq{} because of the recruited moderators' expertise, both in years of experience and in variety of communities moderated.}\chiedit{Although small, the sample was highly experienced: four of five participants had 6--14 years of moderation experience, and participants collectively moderated 10 communities spanning nine topics. We use the study for in-depth examination of experienced practice, not to estimate prevalence across Reddit moderators.}
\begin{table}[t]
\small
\sffamily
\centering
\caption{Summary of participants and the subreddits they moderate. We report the number of years of experience each participant has as a moderator along with the topic and number of subscribers there are in each subreddit they moderate. Subreddits marked with an asterisk (i.e., $\mathcal{S}_1$, $\mathcal{S}_2$, $\mathcal{S}_3$) were utilized to create the corresponding participant's custom \test{}.}
\Description{Summary of participant and subreddits moderated, grouped by participant.}
\label{tab:participants}
\resizebox{\columnwidth}{!}{
\begin{tabular}{ccr@{}llr}
\toprule
\textbf{Participant ID} & 
\textbf{Experience} &
\textbf{Subreddit} &
& 
\textbf{Topic} &
\textbf{Subscribers} \\
\midrule

\multirow{3}{*}{P1} & 
\multirow{3}{*}{12 years} & 
$\mathcal{S}_1$ & 
* &
Sports & 
385K\\ 
 & 
 & 
$\mathcal{S}_4$ & 
 &
Q\&A & 
170K\\ 
 & 
 & 
$\mathcal{S}_5$ & 
 &
Location & 
135K\\ %
\\

\multirow{4}{*}{P2} & 
\multirow{4}{*}{14 years} & 
\textbf{$\mathcal{S}_2$} & 
 * &
Music & 
325K\\ 
 & 
 & 
$\mathcal{S}_6$ & 
 &
History & 
700K\\ %
 & 
 & 
$\mathcal{S}_7$ & 
 &
Transportation & 
10K\\ %
 & 
 & 
$\mathcal{S}_{8}$ & 
 &
Entertainment & 
< 1K\\ %
\\

\multirow{2}{*}{P3} & 
\multirow{2}{*}{13 years} & 
\textbf{$\mathcal{S}_3$} & 
*  &
Sports & 
4M\\ 
 & 
 & 
$\mathcal{S}_9$ & 
 &
School & 
10K\\ %
\\

\multirow{2}{*}{P4} & 
\multirow{2}{*}{6 years} & 
\textbf{$\mathcal{S}_3$} & 
 * &
Sports & 
4M\\ 
 & 
 & 
$\mathcal{S}_{10}$ & 
 &
College Life & 
25K\\ %
\\

P5 & 
< 1 year & 
$\mathcal{S}_2$ & 
 * &
Music & 
325K\\ 

\bottomrule\vspace{-20pt}
\end{tabular}}
\end{table}

\subsection{\chiedit{Setting up a Controlled Study Environment}}
\label{sec:testbed}
For the purpose of the study, we wanted to provide participants with a familiar, but controlled environment to explore the \pq{}. 
For each participant, we collected 105 posts and all their comments from one of the subreddits they moderate, allowing us to populate the \mq{} with multiple pages of familiar content, including both full pages and partial pages.
\edit{Using Python's Reddit API (PRAW) and, for the final three participants, using the Reddit for Researchers API, we collected a sample of non-video posts with fewer than 25 comments for tractability.}

Similar to the approach taken in prior work~\cite{chandrasekharan_crossmod_2019}, we created a private community (denoted by \test{}) on Reddit to serve as a controlled environment to conduct interviews. \test{} is only accessible to the researchers and participants (at the time of their interview). 
We posted a copy of each post and comment collected via PRAW in the same order in which they were originally posted to \test{} using the API. 
This created a subreddit containing 105 posts from the initial two communities. 
For the remaining subreddits, we used Javascript to simulate the HTML for posts and comments in \test{} for the duration of the study.

We manually added three generic flairs to \test{} that would apply to any participant's subreddit: ``Topic Flair,'' ``Format Flair,'' and ``Mod Pick Flair.'' These let moderators use flairs as positive feedback, as some reported doing~\cite{lambert_positive_2024}.

\chicut{The developed browser extension uses JavaScript to manipulate the posts shown to participants during their interview. Each participant only saw posts from one subreddit they moderate to be a convincing mirror of a familiar community. We also manipulate the HTML of the \mq{} to change the author karma, author creation date, and each contributions' score to match the score in the original community at the time of data collection. This increases the believability of \test{} and allows the participant to utilize features related to author characteristics and score. The data collected via the Reddit for Researchers API had anonymized authors, thus we generated synthetic values for each author's account creation date, a field used in the sorting and filtering features. The synthetic data fell on a left-skewed distribution highlighting a peak in account creations around 2020 and correlated with author karma so that older accounts tended to have higher karma (i.e., higher reputation), with some amount of added noise.}\chiedit{Each participant saw content from one community they moderated, with available score and author metadata preserved; synthetic account-age values were used where the researcher API anonymized them.}

\subsubsection{Creating a controlled study environment through \test{}}

\chicut{The usage of a browser extension also allows us to control the actions participants can take during the user study. Specifically, moderation actions like bans, mutes, and the ability to message real Reddit users were disabled to maintain a controlled study environment that has no impact outside \test{}. Similarly, the ``Curate'' and ``Explain'' features only make posts and comments within \test{}. When necessary, functionality was simulated by injecting HTML, guaranteeing no impact beyond \test{}. The approve and remove buttons on the \mq{} were also disabled during the user study so that the same (private) subreddit can be used as a base for \test{} across multiple participants.} \edit{\chicut{Finally, as seen in \autoref{fig:filter_sorting}, the option to sort by ``Most Reported First'' was disabled since there are inherently no reports in \test{}.}}\chiedit{Actions that could reach real users or alter their communities were disabled or simulated.}

\subsection{Setting up the User Study}

\chicut{Each recruited participant engaged in a single one-hour synchronous user study session with the researchers over Zoom. The consent form for participation was shared with each participant by email when scheduling the interview. At the start of each session, the first author shared the consent form with the participant again and allowed for time to ask questions. Once participants verbally consented, the interviewer started recording the meeting and introduced the participant to \test{} as a mirror of a subreddit they moderate that may differ slightly from the original subreddit. They were informed that none of their actions would affect the original subreddit or other Reddit users. Even though participants were shown data from a single subreddit, they were encouraged to provide insights based on moderation practices in any community they moderate. Next, the researchers guided the participant through the process to add the browser extension to their Chrome browser and logging into a Reddit account created for the study. Participants could then to navigate to \test{} and enable the tool. Each participant watched the same pre-recorded demonstration videos for each feature and took a couple of minutes per feature to explore its functionality and ask questions.}\chiedit{Each participant completed a one-hour Zoom session using \test{} as a controlled mirror of one community they moderated, after consent and standardized feature demonstrations.} 

At the end of the feature demonstrations, participants completed two positive-moderation tasks while thinking aloud to surface how they interpreted and combined available signals, decided what warranted encouragement, and composed the available affordances into moderation workflows:

\label{sec:tasks}
\begin{enumerate}
    \item Discovery: Find 2 posts and 2 comments you want to see more of in your community.
    \item Rewarding: For those 2 posts/comments, apply a reward or communicate approval using the new features.
\end{enumerate}

\chiedit{The tasks specified the moderation objective---finding and encouraging contributions---but did not prescribe which signals, criteria, or features participants should use. This allowed us to distinguish task-scaffolded behavior from participant-surfaced practices and appropriations.}

These tasks were selected to uncover how moderators might use the features provided by the \pq{} to reinforce desired behaviors and obtain feature-level feedback.
They also mimic the challenges moderators reported experiencing when employing positive reinforcement~\cite{lambert_positive_2024}.
The participants completed these tasks in tandem by rewarding each post or comment as they discovered it. 
\edit{Participants were instructed to think aloud when they wanted to use a feature that would typically be possible but was disabled as a result of the test environment (e.g., approve, remove, sort by reports, DM users, etc.).}

\subsubsection{Interviews}

Following completion of the tasks, the researchers led a semi-structured interview to collect feedback from the participant about the tool and determine potential improvements. The interviews were audio recorded and anonymized in the transcripts used for analysis.
\chiedit{The transcripts were qualitatively analyzed using an inductive coding approach focused on moderator reasoning and practice: how participants combined signals, decided what to encourage, incorporated or rejected affordances, and surfaced tensions or uses beyond the assigned tasks. We also coded feature-level feedback to understand how particular affordances enabled or constrained these practices.}
Two authors coded one transcript together to be on the same page about the process. Then, each of these authors coded two additional transcripts independently. Each author then reviewed the codes assigned to the two transcripts they did not code and made adjustments as needed to be on the same page about all \nummoderators{} transcripts.
\edit{After the initial interviews, the first author conducted a follow-up coding pass, and a second author independently watched all recordings.
Then, the two authors independently synthesized findings after which they discussed disagreements.}

\subsection{Ethical Considerations}

In designing the user study, we worked closely with the Institutional Review Board (IRB) at the first author's university to ensure protection of the participants through recruitment, interviews, and analysis.
Additionally, in constructing \test{}, we took care to ensure that no actions taken during the study would impact users or contributions externally.
We also guaranteed that all content shown to participants during the study was from a community they moderate to avoid any harms beyond their typical moderation duties.
Participants consented to being interviewed and were assured that they could withdraw at any point. 
All transcripts of the interviews were anonymized to preserve participant privacy.
Additionally, we do not report the specific subreddits moderated by the participants to avoid identification.

\edit{We acknowledge potential biases in the design of the \pq{}. 
Tools that surface content by aggregating group judgments can inherit the subtle
in-group biases those judgments carry~\cite{dovidio_included}.
Because our desirability signal is based on predicted community reception (i.e., predicted \textit{score}), it inherits the known tendency of score-based signals to reflect majority preferences, which may under-surface content valued by smaller segments of a community.
We mitigate this in three ways. First, the desirability signal is one of several discovery metrics rather than the sole input. Second, the \pq{} is model-agnostic and can accommodate richer signals as communities identify what best capture their notion of desirability.
Third, moderators retain full agency over what, if anything, is ultimately rewarded.
Consistent with this, our findings (Section~\ref{sec:AI-reliance}) show that moderators treated the desirability cue as one input among several rather than deferring to it.
}

\section{Findings}
\label{sec:findings}
\chicut{Our findings are organized to surface both feature-level perceptions tied to specific design choices, and workflow-level patterns that an integrated positive moderation tool is uniquely positioned to reveal.}\chiedit{We organize the findings around our research questions, foregrounding moderator practices and appropriations before feature-level feedback. RQ1 examines how moderators combined system-provided signals, realized engagement, and their own judgment. RQ2 examines tensions and additional practices, distinguishing task-scaffolded behavior from uses participants surfaced beyond the assigned discovery-and-reward tasks.}






\subsection{\edit{Workflow Patterns Surfaced by the \pq{}}}

\subsubsection{\edit{Triangulation across AI-driven and community signals}}
\label{sec:AI-reliance}
\edit{
Participants did not use the \chicut{desirability cue}\chiedit{predicted community reception cue} as a primary decision input when choosing content to reward.
Instead, the \chicut{desirability scores}\chiedit{predicted scores} were used as a filtering or sorting criterion to triage potential content to reward.
Two participants used the desirability-related metrics to sort the feed (P4 and P5) and another two used them as filtering criteria (P2 and P3), essentially prioritizing \chicut{higher desirability}\chiedit{higher predicted community reception} when searching for content to reward.
P3 coupled a \chicut{high desirability filter}\chiedit{filter for higher predicted community reception} with sorting by highest karma, which leverages Reddit karma as a signal of credibility within the subset of content highlighted by AI.
Within the triaged set of candidates, participants largely de-prioritized the \chicut{desirability score}\chiedit{predicted community reception} in favor of a post's content, its author's history, and the moderator's own perceptions of ``desirability'' when choosing what to reward.
}

\edit{
We characterize this pattern as triangulation rather than anchoring: \chicut{AI-driven desirability score}\chiedit{predicted community reception} served as confirmation or a flag for additional scrutiny, not as an action trigger.
Behavioral evidence from our user study supports this. 
Of the 13 posts mods identified as reward-worthy, four had ``Undesirable'' or ``Highly Undesirable'' visual indicators, and participant-selected comments were similarly labeled as undesirable by the model about 30\% of the time.
}

\subsubsection{\edit{``Hidden Gems'' Workflows: Prediction vs. Reality}}
\label{sec:hidden-gems}
\edit{Two participants articulated an unanticipated workflow that uses the gap between \chicut{predicted desirability}\chiedit{predicted community reception} and actual community engagement as a discovery signal in its own right.
In these cases, participants treated the model's prediction not as a recommendation but as a counterfactual against which observed community behavior could be compared.
P2 introduced the idea of investigating paradoxes in prediction and reality (e.g., \chicut{high desirability}\chiedit{high predicted reception} but low karma/engagement) as a signal for \inlinequote{undervalued gems}---posts the moderators and/or community \textit{would} find desirable but are not receiving their deserved recognition or attention.
Therefore, P2 asserted that \inlinequote{those paradoxes are probably what require the most attention.}
}

\edit{
Along these lines, r/AskHistorians compiles a weekly thread highlighting overlooked posts. Existing systems can surface content a moderator or curator is likely to approve of~\cite{he_cura_2023,park_supporting_2016,liu_needling_2025}, but neither the \mq{} nor these systems support locating contributions that are likely to be well received but have \textit{not} accumulated engagement---leaving moderation teams to find them by hand.
}

\edit{In practice, the \pq{} could improve the workflow for finding these hidden gems by including bi-directional sorting and filtering metrics that support triaging paradoxical content (e.g., \chicut{high desirability}\chiedit{high predicted reception} but \textit{low} karma or engagement).
Regardless, participants were excited about the capacity for the current Curate feature to specifically support hidden gems.
For example, P1 touted curation as a moderator's \inlinequote{seal of approval, [which] may give the encouragement for other folks to deliver that quality [\dots] of content.}
P1 demonstrated this workflow of using curation to give moderator-favored posts and comments \inlinequote{a second chance at the spotlight.} 
P2 similarly explained:
\blockquote{Maybe somebody made a great post that I think is underappreciated, [\dots] I could add that to the curate list.}{P2}
}

\subsubsection{\edit{Repurposing Positive Features for Punitive Moderation}}
\edit{All interviewed moderators presented ways the positive features could be repurposed to enhance existing punitive workflows. 
Participants noted that sorting by ``most newcomer comments'' may detect brigading (P4 and P5) and sorting by ``newest author first'' may detect spam (P5).
Furthermore, moderators proposed enabling sorting in reverse (e.g., least desirable posts first -- P1 and P2) and allowing the explain feature to communicate why moderators did \textit{not} approve of a post (P3 and P5) to support punitive workflows.
Likewise, P3 suggested integrating private moderator logs into the \chicut{desirability model}\chiedit{community-reception model} to enhance its detection of true ``undesirability.''
 }


\edit{The moderators' desire to integrate the positive and the punitive indicates how porous and ill-defined the boundary is between the two moderation strategies. This porousness is also reflected in P5's description of how AutoModerator---Reddit's automated moderation tool---currently operates in their subreddit. P5 said: \inlinequote{we actually had to turn off the AutoMod, or we cranked down the settings on how aggressive it was, because [\dots] it was putting a bunch of stuff in the queue that wasn't a problem.}
Thus, even tools intended for flagging harmful content for moderator review are not distinguishing between the two classes of content---positive and punishable.
}


\subsubsection{\edit{Retrospective and Auditing Workflows}}
\label{findings:retrospective-audit}
\edit{The \pq{} arms moderators with a unique capability to retroactively audit their community's content.
P3 said \inlinequote{a lot of subreddits do [\dots] a roundup of [\dots] the best comments, the best threads} and highlighted that the discovery features would help moderators identify candidates that have been \inlinequote{most interacted with or that meets certain kind of criteria.}
Both P1 and P3 expressed the utility of the Curate feature in streamlining the manual work of creating weekly roundups.
Furthermore, P4 highlighted how curation extends to more retrospective collections.
They said that \inlinequote{the sorting and filtering options would be useful [\dots] to find older posts that people really liked if we wanted to do a year-end highlight} (P4).
P3 added that the \pq{}'s discovery features could help this retrospective workflow:
\blockquote{You do a filter and a sort [\dots] and all of a sudden, you're only looking for people who are going to qualify for a certain threshold, so that you're not engaging in manipulation.}{P3}
}



\edit{The \pq{} also helps answer questions about what did and did not go well, how to incentivize engagement, and what to promote. 
To this point, P4 explained how the desirability visual cue could help them audit a moderator-created weekly thread which they felt did not have \inlinequote{great traction} in their subreddit. 
During the study, P4 noticed the weekly thread had a ``Neutral'' desirability score, helping them gauge the thread's typical reception from the community.
P3 described the same auditing power the \pq{} gave them and, with respect to audits of recurring posts made by their team, said \inlinequote{currently, if we wanted to do that, because I've done that, we have to do it by hand.}}

\subsubsection{\edit{Connecting to users workflow}}
\label{connecting-to-users}
\edit{Participants surfaced the capacity of the \pq{} to support user outreach.
P1 said: \inlinequote{I don't actually have a good way of knowing who are my contributors to this community,} but thought the sorting menu could help them \inlinequote{find folks that do really good content, and maybe I can identify with them and interact with them in some way. Maybe I reach out to them and ask them to become a moderator.}
The rewarding features too were regarded as useful means \inlinequote{to recognize users} (P4), something they \inlinequote{always kind of think about and want to do, [\dots] any way that we can recognize users for their great contributions is something that we're interested in.}
}

\subsection{\edit{Feature-level Feedback on the \pq{}}}
\edit{This subsection details participant perspectives and suggestions on each new feature with respect to the stated design goals (\autoref{sec:design-goals}).}
\edit{One participant captured the overall reception: the} \pq{} solves \inlinequote{an issue [I have]  with Reddit, which is I don't feel like I have a good way of properly highlighting and acknowledging posts} (P1).
The intuitive and clean UI was highlighted (P2 and P4) along with the smooth integration into the \mq{} (P1 and P4) without disrupting their day-to-day workflows (P1--P5).

\subsubsection{\edit{Discovery is enhanced by improved triaging tools}}
\label{sec:discovery-feedback}
\edit{Participants highlighted how the additional sorting metrics helped them connect to users (P1), the \inlinequote{filters definitely make it faster to find posts} (P3), 
and the two features interact well together:
\blockquote{Reddit does not have great filter options.
    [\dots] It's great to see that there are easy to use filters, but also that you can sort within them, too.}{P4} 
To improve on their functionality, P3 wanted the discovery features to apply to \inlinequote{everything [\dots] within a certain time span,} meaning comments along with posts.
}

\edit{Regarding the desirability cue---particularly the underlying model---participants highlighted how highly-upvoted content might reflect quick-to-consume content favored by the \inlinequote{silent lurking majority} and exclude the opinions of 
users willing to parse the \inlinequote{good content [that] takes longer to engage with} (P2).
P2, P3, and P5 proposed solutions including training the model using the \pq{}'s own features (e.g., curate, award, explain) and giving the model access to a community's \textit{modlogs}---the log of all moderation actions taken---to better capture what content is truly \inlinequote{highly undesirable.}
}
\chiedit{Aggregate reception can privilege content that is quickly or broadly rewarded---engagement bait among the familiar examples---and therefore need not coincide with what moderators judge worth encouraging.}


\subsubsection{\edit{Rewarding features were unanimously favored.}}
\label{sec:flexible-curate}
\edit{The rewarding features received the most positive responses from participants.
In particular, those who used Reddit's native awarding mechanism before the feature underwent changes were interested in our adapted rewards. 
P1 mentioned \inlinequote{the award feature in particular, I would definitely use,} and that it \inlinequote{would replace the `quality content' flair,} an existing way their moderation team provides positive feedback.}

\edit{
The Curate feature received the most interest of any feature in the study. P2 said:
\blockquote{I think the best of the week is fantastic. This is by far the coolest thing.}{P2}}

\edit{With this interest came ideas of how the feature could be more flexible and customizable to integrate into many workflows more seamlessly.
P5 demonstrated how the Curate feature could create a playlist of songs in their music-related community and P1 indicated interest in more categorization options (e.g., memes). This ties closely into P1 and P4's desire to have customizable text in a curated post and variable frequency (e.g., weekly or monthly).}

\subsubsection{\edit{Norm-setting through replies leaves moderators vulnerable to engagement.}}
\label{sec:norm-setting-feedback}


\edit{The Explain feature of the \pq{} elicited many valuable insights on designing ways to communicate positive norms.
On one hand, participants appreciated the default reasons built into the tool. 
P4, for example, said \inlinequote{these explanation reasons reflect that, [\dots] it's nice to see that it [\dots] aligns with our goals for the subreddit.}
}

\edit{Beyond its content, moderators had concerns regarding the mechanism by which explanations are delivered.
P3 stated that the explanation appearing as a comment \inlinequote{could be seen as kind of spammy by users,} P5 said they \inlinequote{don't like a bunch of automated comments gumming up the conversation,} and P2 thought that moderator opinions \inlinequote{throughout the comments would get tiresome for the users.}}

\edit{Importantly, P1 raised concerns beyond user experience, indicating potentially adverse impact on the moderators.
P1 felt that the explanation comment left the moderators  vulnerable by allowing users to \inlinequote{reply directly to the comment} and \inlinequote{downvote the hell out of it.}
As an alternative, P1 suggested displaying the explanation as a \textit{tooltip} that reveals the explanation to users upon hover and prevents interaction:
\inlinequote{It'd almost be like a tooltip [\dots] you would hover over it, it's like, oh, why does this comment have, like, a gold star on it?}
P5 independently suggested displaying a flair or tag attached to the rewarded contribution. Both alternatives prevent interaction from users.
If a moderator \textit{does} want that interaction, 
P3 preferred to send a \inlinequote{private message instead of a public message just so that it doesn't get overly cluttered, but still provides that positive feedback to the individual user.}}


\section{Discussion}
\label{sec:discussion}
\edit{The findings elicited many impressions and perspectives on the \pq{}, offering insights into how designers can best design with moderator workflows and actual interactions in mind.}
This section summarizes the implications of this work for \edit{design and} for moderation more broadly along with future directions for the \pq{}.


\subsection{\edit{Attribution and Channel as Key Design Axes}}
\edit{One key design implication from this study is that delivering positive feedback is a design problem with two axes: \textit{channel} (i.e., how the feedback reaches users) and \textit{attribution} (i.e., whose voice the feedback appears to carry).
Importantly, current positive moderation tooling---including the \pq{}'s Explain feature---does not treat it that way.}
\chiedit{These axes also operate at multiple levels: channel shapes who can see and respond to recognition, while attribution shapes whose judgment that recognition represents. Together, they affect both how recognition is experienced by its recipient and how it is interpreted by the broader community.}

\edit{\subsubsection{Channel:} As discussed in \autoref{sec:norm-setting-feedback}, three participants proposed alternative channels to the \pq{}'s current public-comment delivery for Explain.
This includes a hover tooltip (P1), a flair or tag (P5), and sending a private message (P3), all of which are non-interactive visual indicators rather than comments that users can reply to or downvote. 
Each proposed channel resolves a distinct concern. The tooltip and flair/tag alternatives avoid the parasocial dynamics of a moderator-authored comment that the community can downvote or argue with. 
The DM channel avoids the favoritism complaints that arise when moderator endorsements are publicly visible.
}

\subsubsection{\edit{Attribution}:} 
\label{sec:disc-attribution}

\edit{A common request from participants was more flexibility in deciding who the rewards appear to be coming from. In the \pq{}'s current implementation, rewards are visibly from the moderator's own account.
Some participants wanted attribution to a specific mod account because it distinguishes the action as moderator-issued, in addition to allowing moderators' to personalize the message (P4).
On the other hand, P1 wanted attribution at the team level to avoid favoritism accusations: \inlinequote{this is collectively from the moderators, or even select moderators, like moderators above a certain age, or, only the subreddit owner would be able to do this.}
For large teams, attribution to a dedicated bot or AutoMod was preferred over individual moderators authoring their own Curate threads (P3).}

\edit{From a different angle, P4 noted that even when the moderator account is attributed, the tone of the message may need careful consideration:
\blockquote{I think, especially if we're trying to recognize someone, I think something that's just, like, a little bit more friendly and less sterile. Instead of, `you know, moderators like this post because <XYZ>,' be like `the moderators have found this to be a really great contribution, we want to shout this out for being <XYZ>.' might just be a little bit friendlier and feel less automated and bot-like.}{P4}
}


\edit{Overall, our user study surfaced five attribution options---individual, team, senior-only, AutoMod, bot---and at least four channel options---public reply, tooltip, DM, flair (or tag). 
These two axes determine visibility of feedback to recipients and bystanders (channel) and whether the feedback reads as personal recognition or automated rule output (attribution).
Combining these two axes yields a 5$\times$4 design space and the appropriate cell depends on community context.}\chiedit{ We treat these combinations as a generative set of possibilities surfaced by participants rather than an exhaustive taxonomy of positive-feedback designs.}

\paragraph{\edit{Design Recommendations}}
\edit{Future positive moderation tools should expose attribution and channel as independent, configurable axes rather than committing to one option. 
User-side visibility---who sees the feedback and how it appears---is itself a design axis and should be exposed similarly.}

\subsection{\edit{Supporting Multi-step Moderation Workflows}}
\label{disc:multi-pipelines}
\edit{In accomplishing the positive moderation tasks, the participants largely did not use the features of the \pq{} in isolation. 
Instead, they composed multiple features into multi-stage pipelines.
More precisely, out of the 23 posts and comments participants rewarded during the interview, the most common pipeline participants followed was to filter $\rightarrow$ sort $\rightarrow$ evaluate $\rightarrow$ act.
\chiedit{This sequence was scaffolded by the task, which asked participants to find and then reward content; we therefore treat it as evidence of how moderators compose available affordances, not as an emergent workflow. What the task did not prescribe was which signals participants combined, in what order, or when they stopped trusting them---the dimensions along which participants varied.}
Furthermore, 11 out of 23 were rewarded with not just one, but two different forms of rewards from the \pq{}.}

\edit{In multiple instances, Reddit karma functioned as a credibility floor, desirability scores served as a secondary signal within that filtered subset, and curate/award/explain was the final action. 
P3 set the desirability filter at the midpoint and then sorted the remaining posts by highest karma to produce \inlinequote{a pretty narrow list} of candidates.}
\edit{P5 executed a parallel pipeline during the task itself: sorting by desirability, filtering by score, and then awarding and curating in sequence: \inlinequote{I'll sort by most desirable first, I'll do the filter of score [\dots], I'm a fan of [redacted] so I’ll award that.}}

\edit{\chicut{Importantly, once familiar, the full pipeline is fast.} P3 described how the \pq{} would allow them to put together a curated weekly roundup in \inlinequote{10 minutes.} \chicut{This highlights how the integration cost of adopting positive moderation for moderators not currently doing it is small.}\chiedit{P3's estimate suggests that the \pq{} can reduce friction for a particular task; it does not establish a reduction in moderator labor overall.}}

\paragraph{\edit{Design Recommendations}}
\edit{Positive moderation tools should support interactions between features \chicut{at a low cost to moderators }to enable the multi-step pipelines moderators tend to construct (e.g., between multiple discovery and rewarding features) to make adoption of such tools more plausible for those not currently engaging with positive moderation strategies.}

\subsection{\chiedit{Positive Moderation as Additional Governance Labor}}
\label{sec:labor}
\chiedit{Making an individual action faster is not the same as reducing moderator labor overall. Positive moderation may add recurring work, including finding contributions to recognize, maintaining collections, explaining rewards, auditing interventions, and coordinating across a team. Participants raised this question directly. P1 asked whether they would \inlinequote{actually want a moderator to have that explicit task} and whether to \inlinequote{make time out of my moderation duties.} Repeated positive moderation practices may also create expectations for continued recognition, turning an initially optional activity into recurring governance work.}

\paragraph{\chiedit{Design Recommendations}}
\chiedit{Tools should help moderators decide when this work is worthwhile and make it selective, batchable, and easy to fold into existing routines. Interfaces should also avoid cues that imply comprehensive coverage, such as counts of unrecognized content, which can turn optional recognition into a perceived backlog.}




\subsection{\edit{Configurable Affordances to Support Community Variety}}


\edit{Another design implication is that positive moderation tools should be designed as separable, configurable affordances rather than as a packaged tool with a fixed configuration.}

\edit{First, participants disagreed about how our system should be deployed: some wanted the \pq{} integrated with existing \mq{} feeds (P4), some preferred a separate feed (P1), and one wanted feature-level toggles (P3). \chicut{The disagreement was not random. It tracked community characteristics such as team size, subscriber count, and norms (e.g., P3's preference for feature-level toggles reflected a 30-member mod team's need for granular control, while P5 favored lighter feature use for their contrarian community).}\chiedit{These preferences reflected differences in team size and community norms.}
}

\edit{Second, the Curate feature, which was the most well-received across all participants, was repeatedly requested in more flexible forms (Section~\ref{sec:flexible-curate}).
This included arbitrary themed collections (P1), music playlists (P2), configurable frequency and editable text (P4), and FAQ-style curated lists. 
The key takeaway here is that Curate is not a single workflow, it is an affordance that moderators want to compose into many workflows depending on the context.}

\edit{Third, participants surfaced different preferences for the channel and attribution choices in Explain (Section~\ref{disc:multi-pipelines}). No single fixed implementation can satisfy all of them.}

\edit{These findings suggest that positive moderation is best understood as a set of composable affordances for \textit{discovery} (e.g., filters, sorts, cues), \textit{action} (e.g., award, curate, explain), and \textit{delivery} (e.g., channel, attribution, frequency) that moderators assemble into context-specific workflows.}

\edit{Community variety is the thread that ties these findings together. The \pq{} already supports moderator agency and customizability by limiting automation. Some prior systems automate the process of discovery~\cite{he_cura_2023} while the \pq{} not only empowers moderators to make their own determinations about what they want to discover, but also provides multiple reward options moderators can decide between. Even so, certain communities may require further customizability. P2, for example, asserted that their community members \inlinequote{don't want to hear about what the cops of the subreddit think are good and bad.} 
However, P2 predicted that the system would be better received by communities~\inlinequote{like r/AskScience that are heavily moderated and people have more deference for the experts or the authority.}}

\edit{\chicut{Our findings highlight the importance of customizability and agency to encourage adoption of positive moderation systems: if communities like P2's could enable only the discovery and rewarding features they need, they may be willing to incorporate the system into their moderation.}}

\paragraph{\edit{Design Recommendations}}
\edit{Future positive moderation systems should anticipate variation across communities, make affordances independently configurable, and let moderators tune their parameters (e.g., attribution, channel, threshold, frequency) rather than ship a tool with fixed defaults.}

\subsection{Implications for Moderation}
\chiedit{Taken together, our findings suggest that positive moderation is better understood as recognition infrastructure than as a collection of rewards. Such systems shape what moderators notice, whose judgment becomes visible, and how recognition reaches contributors. In doing so, they allocate attention, authority, and governance labor across a community. Positive moderation is therefore not simply the benign opposite of punitive moderation, but another form of governance through which moderators shape what and whom a community recognizes.}
\subsubsection{\edit{Beyond Gatekeeping: Supporting Community Strategy}}
\edit{
At the start of interviews, we asked each moderator to describe how they normally moderate and whether they reward positive behavior.
Participants explained that they \inlinequote{don't go out of our way to kind of acknowledge anything} (P3) and \inlinequote{don't have any kind of specific system in place} for rewarding (P4), instead focusing on moderating through punitive \inlinequote{auto-moderation tools} (P2), \inlinequote{the comment nuke} (P4), and thresholds that filter \inlinequote{out most of the spam} (P5).}

\edit{\chicut{This policing framing shifted as moderators used the \pq{},} \chiedit{The study also prompted moderators to reflect on community-strategy work beyond enforcement,} particularly when they explored its capacity for retrospective work (Section~\ref{findings:retrospective-audit}). The Discovery and Curation features afforded a kind of community-strategy work that was difficult to do at scale before.
P4 described running surveys to \inlinequote{steer the community towards the discussion that they are looking for,} a community-strategy practice that could be augmented by the \pq{}'s retrospective capacity.}

\edit{\chicut{Through the course of the user study, participants also began articulating the shift in self-perception explicitly.}\chiedit{Participants also articulated strategy-oriented reflections during the study.}
For instance, P1 tried to predict downstream behavioral effects of their positive moderation choices: \inlinequote{if I do curate it, it may increase the desirability of the contents [...] giving it a second chance at the spotlight.} 
P5 reflected on their own framing of moderation during the interview:
\blockquote{I think I do have, like, a negative focus in that I'm not trying to find what's good, I'm trying to find what's bad and get rid of it... maybe that's something that I should fix.}{P5}}

\paragraph{\edit{Design Recommendations}}
\edit{Future positive moderation tools should support both real-time triage and retrospective audit as distinct deployment modes to \chicut{enable moderators to shift toward}\chiedit{support} community-strategy practices not currently facilitated by existing tools, including the \mq{}.}

  



\subsubsection{\chiedit{What Predicted Community Reception Is, and Is Not}}
\chiedit{The \pq{}'s desirability cue surfaces predicted community reception: model scores estimating how a subreddit would receive a contribution.}
\chiedit{Reddit exposes no moderator-only endorsement signal---a moderator's upvote is indistinguishable from any other---so whole-community reception provides an observable community-level signal for the upvoting behavior moderators report using to reinforce contributions they value~\cite{lambert_positive_2024}, rather than a direct measure of moderator preference. Two distinctions matter. First, reception here is predicted, not observed: potentially well-received contributions can go unseen~\cite{gilbert_widespread_2013}, so the cue is not redundant with ranking content that has already accumulated engagement. Participants used this exact prediction--engagement mismatch to identify overlooked contributions (\autoref{sec:hidden-gems}). Second, predicted community reception is not moderator endorsement. Their divergence in our study is an empirical result about how moderators combine community-level signals with their own judgment, not a defect of the signal.}

Some of the participants mentioned existing concerns with fairness in their moderation practices, especially when trying to incorporate positive feedback.
\chicut{At the same time, the participants failed to recognize potential differences in the content they and their communities find desirable. For example, moderators typically blamed the desirability model when it scored a contribution differently than they would have. Even after clarification that the model estimates the degree to which their community would score a contribution, the participants did not acknowledge that unexpected model scores may have something to do with differences between how moderators and community members view content.}
\chiedit{When predictions differed from their judgments, participants often attributed the mismatch to model error rather than to possible moderator--community differences.}

\chicut{The participants' willingness to accept desirability scores aligning with their mental model and unwillingness to accept those that did not indicates that users of systems like the \pq{} may not be overreliant on AI.}
\chiedit{Participants tended to accept predictions that matched their expectations and question those that did not, consistent with AI serving as confirmation rather than authority,} an established behavior in understanding human interaction with AI~\cite{sharma_generative_2024, rosbach_when_2025, yin_does_2024, du_confirmation_2025, bashkirova_confirmation_2024}.
This pattern indicates a need for moderators to determine any misalignment between \chicut{user and moderator definitions of desirability}\chiedit{broader community reception and moderator preferences}, similar to prior work exploring user-moderator alignment with respect to community rules~\cite{koshy_measuring_2023,koshy_venire_2025}.
One participant even mentioned using surveys to help the moderation team \inlinequote{steer the community towards the discussion that they are looking for} (P4).


\paragraph{\edit{Design Recommendations}}
\edit{\chicut{Designers of positive moderation tools should empower moderators to align their definitions of desirability with those of their communities, in part by surfacing community reception alongside AI predictions and enabling periodic recalibration so moderation teams can maintain alignment as norms evolve.}\chiedit{Designers should make predicted reception, realized engagement, and moderator endorsement visibly distinct and support comparison among them.}}

\subsection{Limitations and Future Work}
\subsubsection{Mitigating the Effects of ``the Rich Get Richer.''}
The design of the \pq{} has the potential to enable moderators to enact the idiom ``the rich get richer.''
More precisely, the tool introduces features for moderators to easily find content that has already received community approval or was posted by an author with already high karma.
Moderators can then apply additional rewards to the same users and contributions that are rich with positive feedback.
However, the \pq{} intentionally integrates newcomers in its design which can help draw moderator attention to users who would especially benefit from positive feedback.
Future work can explore more explicitly how positive feedback may contribute to ``the rich get richer'' idiom in practice.





\subsubsection{\edit{Sample Diversity, Deployment, and Community-level Effects}}

\edit{This work is limited by the number of moderators who participated and by the controlled, one-hour user study format. We cannot make claims about the long-term usage of the \pq{} or its impact on a community.
Our findings surface how experienced moderators would use such an integrated positive moderation system, not measurements of effect.
\chicut{That said, our participants had significant expertise and variety in moderation practices, so we believe the sample is sufficient to gather perceptions and feedback on the design goals.}
\chiedit{The study draws on substantial practitioner expertise: four of five participants had 6--14 years of experience and collectively moderated 10 communities across nine topics. This supports in-depth examination of practice, while limiting the breadth and prevalence claims we can make.}
Our sample nonetheless skewed toward moderators of broadly-themed communities (e.g., sports, music, Q\&A). Moderators of more sensitive or \chicut{identify-based}\chiedit{identity-based} communities might surface additional insights about when positive moderation helps or harms.}

\edit{\chicut{The small sample also reflects the difficulty of recruiting moderators for research. Across four waves of outreach to 49 moderation teams comprising more than 2{,}000 moderators, only five completed our study, a response rate of roughly 0.25\%. Such low response rates are a structural constraint on studies of moderators and other hard-to-reach populations, placing a practical ceiling on sample size and diversity. One promising complement is agent-based simulation of the moderation process.}}
\chiedit{Volunteer moderators are difficult to recruit, and our four-wave direct-modmail recruitment followed methods established in prior Reddit moderation research~\cite{song_modsandbox_2023,koshy_venire_2025,bajpai_think_2025,liu_needling_2025} (\autoref{sec:recruitment}).}
\edit{A natural next step\chiedit{---one that would depend on the sustained community relationships such recruitment builds toward---}is a longer-term deployment of the \pq{} in a small number of communities, measuring outcomes such as changes in the behavior of users who receive positive feedback, changes in moderator-reported workload, and the rate and composition of encouraged or curated content over time.}

\subsubsection{\chicut{Expand modeling of community-specific desirability}\chiedit{Expand modeling of community-specific reception and moderator preference}}
\edit{Our desirability cue predicts community reception (score), which does not always match the quality a moderator wants to reinforce---a gap our study surfaces as a finding rather than a defect (\autoref{sec:AI-reliance}). \chicut{This points to several ways to improve the model in future work (\autoref{sec:discovery-feedback}).}\chiedit{Future systems can model whole-community reception and moderator preference as related but distinct signals.}}
\chicut{Firstly, the model itself would benefit from validation from human raters, particularly those from corresponding communities, to examine whether the model's predictions align with community definitions of desirability.}\chiedit{Firstly, participants proposed using \pq{} actions and moderation logs as inputs to a separate moderator-specific model alongside whole-community reception.}

Secondly, the model was intended to predict the likelihood that a community would respond well to a comment or post (i.e., lead to a high score).
This means that a contribution marked as ``Highly Undesirable'' is likely not to receive many upvotes, but does not necessarily contain norm-violating behavior.
\chicut{However, the modeling, as indicated by several participants, would benefit greatly from direct community input. In particular, a desirability model that incorporates attributes of frequently removed contributions would be more capable of approximating not only what the community might find desirable, but also what the moderation team is looking to reward.}

\edit{Thirdly, score is an anonymous signal of community reception: it aggregates upvotes without revealing whether they come from a moderator or from other community members.
Although moderators commonly rely on upvotes to reward desirable behavior~\cite{lambert_positive_2024}, a moderator's upvote carries no more visible weight than any other, which limits score's usefulness as a deliberate signal of endorsement. This points to a broader limitation of anonymous signals like score, which cannot reveal who endorsed a contribution. Moderator-specific signals of positive feedback---like YouTube's Creator Hearts~\cite{choi_creator_2024} or the New York Times' ``NYT Picks''~\cite{wang_highlighting_2021}---make that source legible, and designing signals that carry moderator attribution while preserving community-level reception (\autoref{sec:disc-attribution}) is a promising direction for future positive moderation tools.}

\section{Conclusion}
Reddit itself has been making changes recently to interface features moderators use to
highlight high-quality content, namely pinning.\footnote{\url{https://www.reddit.com/r/modnews/comments/1cnacle/comment/l62tqrw/}}
Reddit's apparent interest in improving highlighting mechanisms speaks to the immediate
importance of designing tools for positive feedback. This work presents the \pq{}, a
positive moderation system built into Reddit's \mq{} that provides a dedicated space to
discover and reinforce contributions moderators want to encourage. In a user study with
\nummoderators{} experienced Reddit moderators, we examine how moderators operationalize
positive reinforcement. Our findings show how moderators combine community signals with
their own judgment, identify overlooked contributions, and repurpose positive affordances
for other moderation work. Together, these findings motivate a view of positive moderation
as \textit{recognition infrastructure}: systems that shape what moderators notice, whose
judgment becomes visible, and how recognition reaches contributors. We derive design
considerations for building such infrastructure around community reception, attribution,
labor, and community fit.

\bibliographystyle{ACM-Reference-Format}
\bibliography{bib}


\end{document}
\endinput